\documentclass[11pt,a4paper]{article}
\usepackage{jheppub}
\usepackage{graphicx}
\usepackage{amsmath}
\usepackage{amssymb}
\usepackage{amsfonts}
\usepackage{hhline}  
\usepackage{url}
\usepackage{caption} 
\usepackage{epsfig}
\usepackage{mathrsfs}
\usepackage{multirow}
\usepackage{physics}
\usepackage{color}
\usepackage{hyperref}
\usepackage{enumitem}
\usepackage{outlines}
\usepackage{multirow}
\usepackage{makecell}
\usepackage{subcaption}

\allowdisplaybreaks
\usepackage{tikz}

\abstract{We propose an idea to build a bridge between reheating and late-time observations in quintessential inflation  by backtracking the evolution of the inflaton field from the present time to the end of reheating. This idea is implemented when the potential gradient is negligible compared to the Hubble friction, rendering the inflaton field frozen, till the present time. We find a simple analytic relation between the reheating temperature and the observational parameters for dark energy, and numerically confirm its validity for typical models of quintessential inflation. This relation is universal and can apply to all quintessential inflation models with any reheating mechanism. It also implies that any quintessential inflation model with a successful reheating with the reheating temperature $1\textrm{MeV}\lesssim T_\textrm{re}\lesssim 10^{15}\textrm{GeV}$ predicts the equation of state of dark energy today extremely close to $-1$, i.e. $-1+10^{-60}\lesssim w_0\lesssim -1+10^{-24}$, unless the inflaton field unfreezes before the present time.}

\title{Bridging between reheating and late-time observations in quintessential inflation}

\author{Ok Song An,\,\,}
\emailAdd{os.an@ryongnamsan.edu.kp}
\author{Jin U Kang,\,\,}
\emailAdd{ju.kang0718@ryongnamsan.edu.kp}
\author{Yong Jin Kim }
\emailAdd{cioc6@ryongnamsan.edu.kp}
\author{and\, Ui Ri Mun\,\,}
\emailAdd{ur.mun0826@ryongnamsan.edu.kp}

\affiliation{Department of Physics, \textbf{Kim Il Sung} University, Ryongnam Dong, Taesong District, Pyongyang, Democratic People's Republic of Korea}

\begin{document}
	\maketitle
\section{Introduction}\label{sec-intro}
The unexpected discovery of an accelerated expansion of our universe in 1998 \cite{SupernovaSearchTeam:1998fmf,SupernovaCosmologyProject:1998vns} raised one of the biggest questions to human beings as to the predominant component of the current universe called \textit{dark energy}.  Cosmic observations \cite{planckR} indicate that dark energy comprises about $69\%$ of the universe and has the equation of state very close to $-1$ at the present time.
Among dark energy candidates, one of the most well-known ones is quintessence  \cite{Zlatev:1998tr,Peebles:2002gy,PhysRevD.37.3406,Steinhardt:1999nw,Armendariz-Picon:2000nqq}, a canonical scalar field minimally coupled to gravity which currently has the potential energy significantly greater than the kinetic energy. Quintessence can successfully avoid the fine tuning problem of the cosmological constant, but it suffers from another tuning problem of its initial conditions. One way to evade this problem is to attribute the early and late-time accelerated expansions to dynamics of the same scalar field so that the initial conditions of quintessence are given by the inflationary attractor which is strongly supported by observations \cite{Planck:2018jri}. This unified approach 
  called \textit{quintessential inflation} \cite{Peebles:1998qn} has  drawn a lot of attention \cite{Peloso:1999dm,Piedipalumbo:2011bj,deHaro:2016cdm,Kleidis:2019ywv,Benisty:2020vvm,Benisty:2020xqm,Yahiro:2001uh,Sen:2000ym,Haro:2019peq} (see also \cite{deHaro:2021swo} for a review and references therein).

  A price to pay for the theoretical elegance of quintessential inflation is to introduce a potential which does not possess a local minimum so that the inflaton does not oscillate to decay into other particles and survives until today. This makes the reheating mechanisms in quintessential inflation quite different from those in standard inflation.   
 Typical reheating mechanisms in quintessential inflation include the gravitational particle production \cite{Ford:1986sy,deHaro:2019oki,Salo:2021vdv,Salo:2021piz,Haque:2022kez,Chun:2009yu}, the instant preheating \cite{Felder:1998vq,Dimopoulos:2017tud} and the curvaton reheating \cite{BuenoSanchez:2007jxm,Feng:2002nb,Agarwal:2018obp}. 
 
 From the cosmological point of view, the most important macroscopic parameters that characterize the reheating period are the reheating temperature $T_\textrm{re}$ and the equation of state parameter $w_\textrm{re}$ during reheating.\footnote{Another macroscopic parameter related to the reheating is the number of e-folds $N_\textrm{re}$, which characterizes how long this period lasted. As we will show explicitly (see eq. \eqref{Nre}), $N_\textrm{re}$ can be fixed for a given inflationary model if $T_\textrm{re}$ or $w_\textrm{re}$ is known.} Gravitational wave background can in principle be used as a direct probe to the reheating temperature \cite{Ghiglieri:2020mhm,Ringwald:2020ist}, however in practice it turns out to be very challenging \cite{Drewes:2024}. 
 It is possible to constrain the reheating temperature indirectly from the CMB observations \cite{Martin:2010kz,Cook:2015vqa,Ueno:2016dim,DiMarco:2017zek,German:2020cbw,Ellis:2021kad,Drewes:2017fmn,Drewes:20242}, provided that $w_\textrm{re}$ is specified. The rigorous determination of $w_\textrm{re}$ calls for the calculation of the dissipation rate of the inflaton (see e.g. \cite{Drewes:2013iaa,Buldgen:2019dus,Cheung:2015iqa}) and in general involves numerical studies \cite{Munoz:2014eqa,Antusch:2020iyq,Saha:2020bis,Lozanov:2017hjm,Lozanov:2016hid} due to great complexity of the reheating phase.  There are few exceptions such as the perturbative reheating by a polynomial potential \cite{Mukhanov:2005sc} and the reheating in the quintessential inflation \cite{deHaro:2023} in which the value of $w_\textrm{re}$ can be assumed a priori from physical intuitions. 
 In particular, it was shown in \cite{deHaro:2023} that the Planck results \cite{Planck:2018jri} do not give a stringent constraint on the reheating temperature in quintessential inflation.\footnote{For determining the reheating temperature microphysically in quintessence inflation, see \cite{deHaro:2022ukj,AresteSalo:2017lkv,AresteSalo:2021lmp,DeHaro:2017abf}.}
 All in all, the reheating still remains the most unconstrained period from observations, because of the lack of knowledge about the microphysical processes governing this period and of the shortage of measurements possibly sensitive to it.
 
 Now we note that quintessential inflation models should obey not only the CMB constraints on inflation but also late-time observational constraints on dark energy irrelevant for standard inflation models. Since quintessential inflation models purport to cover the cosmic phases from inflation through reheating to late-time accelerating era, then it is natural to think that reheating parameters should be linked to  the late-time observations apart from the inflationary imprints on CMB. The main goal of the present work is to explore this link quantitatively and provide the best information available from it.  
 The idea we employ is to backtrack the evolution of the inflaton field from today to the end of reheating with the final conditions at the present time constrained by the observations on dark energy, instead of tracking from the end of inflation forward in time with initial conditions at early times set by hand.
 
 By inspecting the dynamics of the inflaton field, one can understand that soon after reheating the inflaton field almost freezes at some model-dependent value due to large Hubble friction until the potential gradient becomes large enough to thaw it (see \cite{Dimopoulos:2017zvq} for a detailed discussion). 
 The important point is that the current observations do not tell us whether or not  the inflaton field is frozen \textit{today}. Namely there are to  two possibilities regarding the behavior of the inflaton field today:  (i) it still remains frozen; (ii) it starts to feel the potential gradient and roll very recently.\footnote{If the inflaton field unfreezes too early, the current value of the equation of state of dark energy would deviate too much from $-1$ to be compatible with observation.} When restricted to quintessential behavior, the first case corresponds to a freezing model  and the second case to a thawing model  \cite{0505494}. In this work, we suppose that the inflaton field is frozen at the present time. In fact, the observed value of the equation of state of dark energy being very close to $-1$ is easily explained when the inflaton field is almost frozen today.\footnote{Quintessential inflation models can be distinguished from $\Lambda\textrm{CDM}$ as the accuracy of CMB observations increases, given that they predict the scalar spectral index slightly greater than that in standard inflation models \cite{Akrami:2017cir}.}  
 
With these motivations in mind we focus on the freezing case and explore the interrelation between reheating and late-time observations. In particular, we find that a successful reheating  with the reheating temperature  $1\textrm{MeV}\lesssim T_\textrm{re}\lesssim 10^{15}\textrm{GeV}$ implies the equation of state of dark energy today extremely close to $-1$, namely $-1+10^{-60}\lesssim w_0\lesssim -1+10^{-24}$.
 
  Our paper is organized as follows. In section \ref{sec-MI consideration} we first review the approach widely adopted to standard inflation models to investigate the interplay between inflation and reheating. Then we describe a novel idea of backtracking to build a bridge between reheating and late-time observations, based on which we find a simple analytic relation between the reheating parameters and late-time observational parameters, without assuming a specific reheating mechanism.   
  This idea is numerically implemented for typical models of quintessential inflation in section \ref{sec-examples}, where we demonstrate the robustness of the relation between the reheating parameters and the late-time observational parameters under the physically viable evolution of the universe after reheating. Finally, we conclude in section \ref{sec-conclusions}.

Throughout this paper we use Planckian units $c=\hbar=k_B=\frac{1}{8\pi G}=1$, unless otherwise specified.

\section{General consideration on reheating and cosmic observations}\label{sec-MI consideration}
\subsection{Reheating and inflationary predictions}

\begin{figure}[!htb]
	\centering
	\begin{tikzpicture}[scale=0.9]
	\draw[->,thick] (0,0)--(0,7) node [anchor=east,color=black] {$\ln{(aH)^{-1}}$};
	\draw[->,thick] (0,0)--(14.3,0) node [anchor=north,color=black] {$ \ln a $};		
	\draw (1,5.1)--(13.5,5.1);
	\node at (0.8,5.4) {$k_*$};
	\draw (1.1,6)--(5.9,1.2);
	\draw (5.9,1.2)--(7.5,1.8);
	\draw (7.5,1.8)--(11.4,5.7);
	\draw (11.4,5.7)--(12.4,6.2);
	\draw (12.4,6.2)--(13,5.6);
	\draw [dashed](2,5.1)--(2,0);
	\node at (2,-0.3) {$a_k$};
	\draw [dashed] (5.9,1.2)--(5.9,0);
	\node at (5.9,-0.3) {$a_\textrm{end}$};
	\draw [dashed] (7.5,1.8)--(7.5,0);
	\node at (7.5,-0.3) {$a_\textrm{re}$};
	\draw [dashed] (13,5.6)--(13,0);
	\node at (13,-0.3) {$a_0$};
	\draw [dashed] (11.4,5.7)--(11.4,0);
	\node at (11.4,-0.3) {$a_\textrm{eq}$};
	\node at (3.2,3) {Inflation};
	\node at (7,1) {Reheating};
	\node at (11.6,6.2) {MD};
	\node at (10,3.6)  {RD};
	\node at (13,6.2)  {$\Lambda$D};
	\node at (2.2,5.4)  [color=blue] {P};
	\node at (6,1.5) [color=blue] {E};
	\node at (7.4,2) [color=blue] {R};
	\node at (11.1,5.8) [color=blue] {Q};
	\end{tikzpicture}
	\caption{Evolution of the comoving Hubble horizon in different epochs of cosmic expansion. RD and MD respectively indicate the radiation and matter dominated eras and $\Lambda\textrm{D}$ the late-time dark energy dominated era.  We denoted the comoving wave number corresponding to a pivot scale by $k_*$  and the scale factors at the horizon crossing of the pivot scale, the end of inflation, the end of reheating, the matter-radiation equality and present time by $a_k$, $a_\textrm{end}$, $a_\textrm{re}$, $a_\textrm{eq}$ and $a_0$, respectively. The points P, E, R and Q correspond to the horizon crossing, the end of inflation, the end of reheating and the matter-radiation equality. The gradients of the lines were chosen from the equation of state parameter of a single component which dominates the energy density of the universe at each era.} \label{fig_aH}
\end{figure}
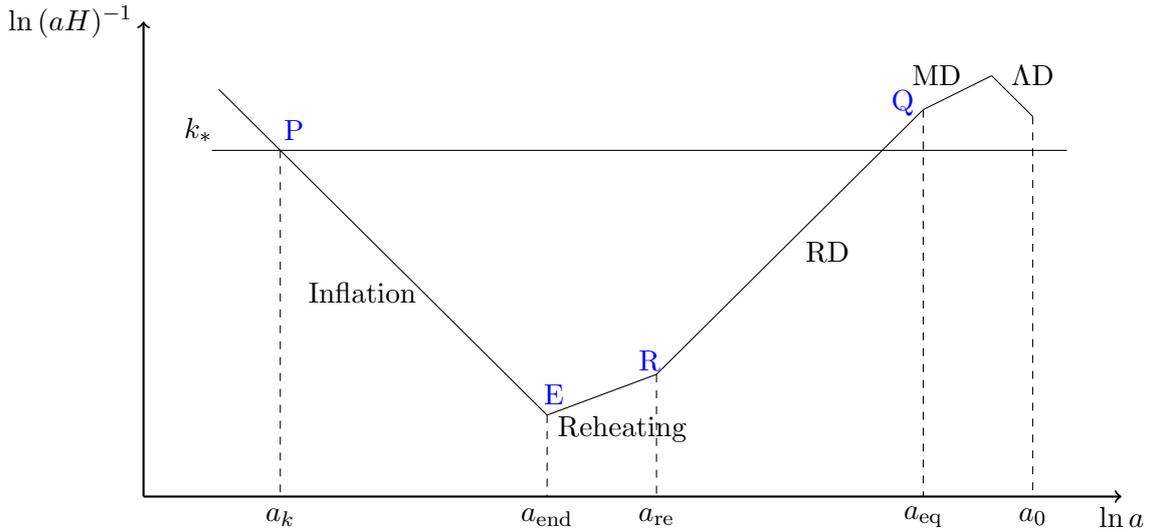

We briefly review an approach widely used in literature \cite{Martin:2010kz,Cook:2015vqa,Ueno:2016dim,DiMarco:2017zek,Drewes:2017fmn,German:2020cbw,Ellis:2021kad,Drewes:20242} to constrain the reheating period by using observational parameters predicted by inflation. 
Figure \ref{fig_aH} depicts the evolution of comoving Hubble horizon in log-log plot\footnote{In figure \ref{fig_aH} we presume that the universe underwent radiation domination (RD), matter domination (MD) and dark energy domination ($\Lambda$D) after the end of reheating. There are alternative scenarios in which the reheating was followed by MD due to heavy particles that decay into radiation later on, see e.g. \cite{deHaro:202310}.}, which  helps us to manifest the uncertainties inherent in the reheating period although the parameters of the inflaton potential are completely known.

Note first in figure \ref{fig_aH} that point Q corresponds to the matter-radiation equality and can be fixed independently of the inflaton potential, since we know the scale factor $a_\textrm{eq}$ and the Hubble parameter $H_\textrm{eq}$ at the matter-radiation equality from observations. Second,  we can determine the scale factor $a_k$ and the Hubble parameter $H_k$ at horizon crossing of the pivot scale with the comoving wave number $k_*$ for a given inflationary model. This further allows us to compute the scale factor $a_\textrm{end}$ and the Hubble parameter $H_\textrm{end}$ at the end of inflation and so fix points P and E. 
However, point R corresponding to the end of reheating can move along the radiation line, since we do know neither the slope of the reheating line, namely the equation of state parameter of reheating $w_\textrm{re}$ nor the reheating temperature $T_\textrm{re}$. Unless we specify the reheating mechanism, we can not determine $T_\textrm{re}$ and $w_\textrm{re}$. Once one of these two parameters is known somehow, we can fix point R, in other words, the scale factor $a_\textrm{re}$ and the Hubble parameter $H_\textrm{re}$ at the end of reheating, which enables us to determine the other reheating parameters including the number of e-folds during reheating $N_\textrm{re}$. So we can conclude that the uncertainties related to the reheating can be included in $T_\textrm{re}$ or $w_\textrm{re}$.

To be more precise, let us express $N_\textrm{re}$ and $T_\textrm{re}$ in terms of CMB observables, model parameters and $w_\textrm{re}$. For the detailed derivation we refer to \cite{Drewes:2017fmn}. We start from the well known relation
\begin{align}\label{aH-k}
0=\ln\left(\frac{k_*}{a_kH_k}\right)=\ln\left(\frac{a_\textrm{end}}{a_k}\frac{a_\textrm{re}}{a_\textrm{end}}\frac{a_0}{a_\textrm{re}}\frac{k_*}{a_0H_k}\right)=N_k+N_\textrm{re}+\ln\left(\frac{a_0}{a_\textrm{re}}\right)+\ln\left(\frac{k_*}{a_0H_k}\right).
\end{align}
 Here the number of e-folds $N_k$ between the horizon crossing and the end of inflation is given by
 \begin{equation}
N_k=\ln\left(\frac{a_\textrm{end}}{a_k}\right)\simeq\int_{\varphi_\textrm{end}}^{\varphi_k} d\varphi \frac{V(\varphi)}{V_{\varphi}}, \label{N_k}
 \end{equation}
where $V(\varphi)$ is the inflaton potential and $V_\varphi\equiv\frac{dV}{d\varphi}$. The values of the inflaton field $\varphi_k$ and $\varphi_\textrm{end}$ at the horizon crossing and the end of inflation can be expressed in terms of model parameters and the amplitude of scalar perturbations $A_s$ by inverting
\begin{equation}\label{Hk-phik}
	H_k=\frac{\pi\sqrt{rA_s}}{\sqrt{2}}=\pi \sqrt{8A_s\epsilon(\varphi_k)}\simeq \sqrt{\frac{V(\varphi_k)}{3}},
\end{equation}
\begin{equation}\label{phi_end}
	\epsilon(\varphi_\textrm{end})=1,
\end{equation}
where $r$  is the tensor-to-scalar ratio and we used the consistency relation $r=16\epsilon(\varphi_k)$. The potential slow-roll parameters $\epsilon$ and $\eta$ are defined as
\begin{equation}\label{slowroll parameter}
\epsilon(\varphi)=\frac{1}{2}\left(\frac{V_{\varphi}}{V}\right)^2,\, \eta(\varphi)=\frac{V_{\varphi\varphi}}{V}.
\end{equation}
Assuming the entropy conservation after the reheating, namely\footnote{Here we have assumed that the effective number of neutrino species $N_\textrm{eff}$ is three, given that $N_\textrm{eff}=3.02\pm 0.17$ \cite{Planck:2018jri}.}
\begin{equation}\label{Tre-T0}
	g_{s,\textrm{re}}a_\textrm{re}^3T_\textrm{re}^3=\frac{43}{11}a_0^3T_0^3,
\end{equation}
with $g_{s,\textrm{re}}$ the effective number of degrees of freedom in entropy at the end of reheating and having the radiation energy density at the end of reheating satisfying
\begin{equation}\label{rho_re}
	\rho_\textrm{re}=\frac{\pi^2g_{*,\textrm{re}}}{30}T_\textrm{re}^4=\rho_\textrm{end} e^{-3(1+w_\textrm{re})N_\textrm{re}}=\frac{4}{3}V_\textrm{end}e^{-3(1+w_\textrm{re})N_\textrm{re}},
\end{equation}
with $g_{*,\textrm{re}}$ the effective number of relativistic degrees of freedom at $T_\textrm{re}$ and $V_\textrm{end}\equiv V(\varphi_\textrm{end})$, we find
\begin{equation}\label{are-a0}
\ln\left(\frac{a_\textrm{re}}{a_0}\right)=\frac{1}{3}\ln\left(\frac{43}{11g_{s,\textrm{re}}}\right)+\frac{1}{4}\ln\left(\frac{\pi^2g_{*,\textrm{re}}}{30}\right)+\frac{1}{4}\ln\left(\frac{3T_0^4}{4V_\textrm{end}}\right)+\frac{3N_\textrm{re}(1+w_\textrm{re})}{4}.
\end{equation}
Plugging eq. \eqref{are-a0} back into eq. \eqref{aH-k}, we obtain
\begin{align}
N_\textrm{re}&=\frac{4}{3w_\textrm{re}-1}\left[N_k+\ln\left(\frac{k_*}{a_0T_0}\right)+\frac{1}{4}\ln\left(\frac{40}{\pi^2g_{*,\textrm{re}}}\right)+\frac{1}{3}\ln\left(\frac{11g_{s,\textrm{re}}}{43}\right)-\frac{1}{2}\ln\left(\frac{16\pi^2\epsilon(\varphi_k)A_s}{2V_\textrm{end}^{1/2}}\right)\right]\nonumber\\
&\equiv \frac{4}{3w_\textrm{re}-1} \mathcal{N}(\mathcal{P}_i,\mathcal{O}_i).\label{Nre}
\end{align}
Here we have introduced a quantity $\mathcal{N}(\mathcal{P}_i,\mathcal{O}_i)$ that can be determined by model parameters $\mathcal{P}_i$ appearing in the inflaton potential and observational parameters $\mathcal{O}_i$ such as the amplitude of scalar perturbations and the scalar spectral index. Note that $\mathcal{O}_i$ represent mostly inflationary parameters except for $T_0$, the CMB temperature today.
We can also express the reheating temperature by using eqs. \eqref{rho_re} and \eqref{Nre}
\begin{equation}
	T_\textrm{re}=\left(\frac{40V_\textrm{end}}{\pi^2g_{*,\textrm{re}}}\right)^{1/4}\exp\left[\frac{3(1+w_\textrm{re})}{1-3w_\textrm{re}}\mathcal{N}(\mathcal{P}_i,\mathcal{O}_i)\right].\label{Tre}
\end{equation}
Eqs. \eqref{Nre} and \eqref{Tre} explicitly show that the e-folding number $N_\textrm{re}$ and the reheating temperature $T_\textrm{re}$ for a given model of inflation can be determined only when the equation of state parameter $w_\textrm{re}$ during reheating is specified. Conversely, if $T_\textrm{re}$ is known, the other two parameters $w_\textrm{re}$ and $N_\textrm{re}$ can be determined 
for a given model of inflation.

Determining $w_\textrm{re}$ or $T_\textrm{re}$ in general depends on reheating mechanisms, which calls for the knowledge of microphysics that controls the dissipation of the inflaton energy into radiation and often entails involved numeric calculations. However, given that the quintessential inflation is followed by kination during which the kinetic energy of the inflaton field dominates, we have  $w_\textrm{re}=1$ which gives rise  the reheating temperature 
\begin{equation}
T_\textrm{re}=\left(\frac{40V_\textrm{end}}{\pi^2g_{*,\textrm{re}}}\right)^{1/4}\exp\left[-3\mathcal{N}(\mathcal{P}_i,\mathcal{O}_i)\right].\label{Tre-QI}
\end{equation}
This shows that the reheating temperature can be determined by model parameters
and observational parameters regardless of reheating mechanisms. But,
as it was demonstrated in \cite{deHaro:2023}, eq. \eqref{Tre-QI} could not provide a stringent constraint on $T_\textrm{re}$ due to the lack of precision of observational constraints on cosmological parameters, e.g. the scalar spectral index. 

The bottom line is that either the equation of state parameter $w_\textrm{re}$ or the reheating temperature $T_\textrm{re}$ still remains a free parameter so that the reheating  becomes the dark period in the inflationary cosmology until we completely understand the reheating mechanisms.  

\subsection{Reheating and late-time observations}\label{sec-rehating and LTO}
From its very definition, any quintessence inflationary model should also be in good agreement with observations related to the late-time accelerated expansion of the universe, which are irrelevant for standard inflation. 
More precisely, the energy density $\rho_{\varphi 0}$ and equation of state parameter of the inflaton field at present time $t_0$ are constrained by 
\begin{equation}
\rho_{\varphi 0}=3\Omega_\Lambda H_0^2,\label{BC-1}
\end{equation}
\begin{equation}
	\frac{\frac{\dot{\varphi}^2(t_0)}{2}-V(\varphi_0)}{\frac{\dot{\varphi}^2(t_0)}{2}+V(\varphi_0)}=w_0,\label{BC-2}
\end{equation}
with $\varphi_0\equiv\varphi(t_0)$ and the current values of the density parameter and the equation of state  for dark energy respectively given by $\Omega_\Lambda=0.685\pm 0.007$ and $w_0=-1.028\pm 0.032$ \cite{planckR}. In what follows we demonstrate how these constraints can help us to build a bridge between reheating and late-time observations.

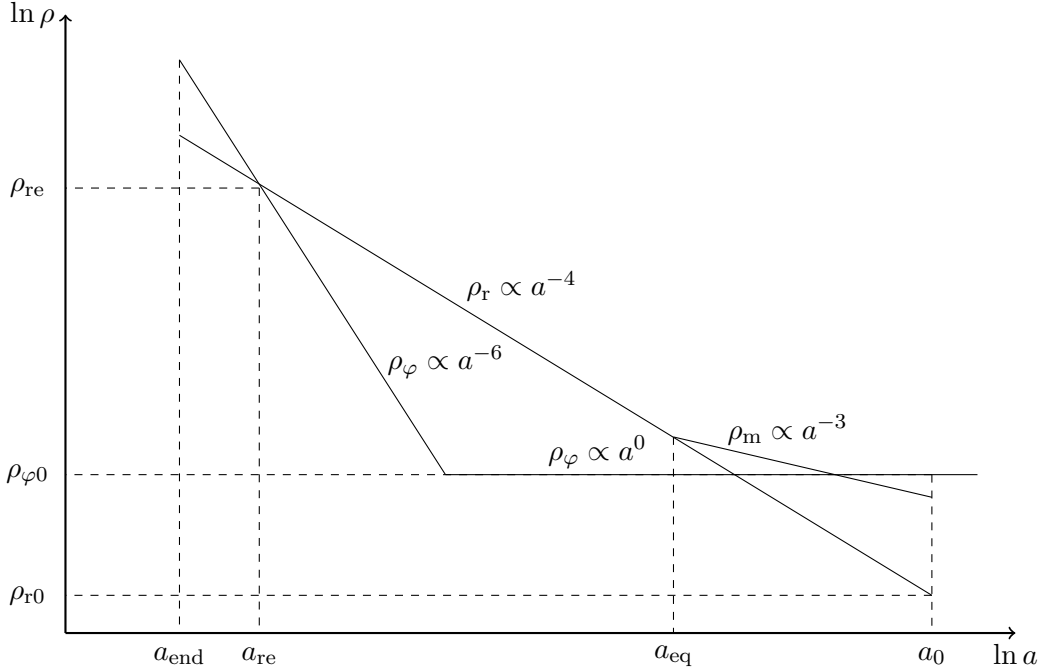
\begin{figure}[!htb]
	\centering
	\begin{tikzpicture}[scale=1]
	\draw[->,thick] (0,-1.6)--(0,6.6) node [anchor=east,color=black] {$\ln{\rho}$};
	\draw[->,thick] (0,-1.6)--(12.5,-1.6) node [anchor=north,color=black] {$ \ln a $};		
	\draw (1.5,6)--(5,0.5);
	\draw (1.5,5)--(11.4,-1.1);
	\draw (8,1)--(11.4,0.2);
	\draw (5,0.5)--(12,0.5);
	\draw [dashed] (1.5,6)--(1.5,-1.6);
	\node at (1.5,-1.9) {$a_\textrm{end}$};
	\draw [dashed] (2.55,4.3)--(2.55,-1.6);
	\node at (2.55,-1.9) {$a_\textrm{re}$};
	\draw [dashed] (8,1)--(8,-1.6);
	\node at (8,-1.9) {$a_\textrm{eq}$};
	\draw [dashed] (11.4,0.5)--(11.4,-1.6);
	\node at (11.4,-1.9) {$a_0$};
	\draw [dashed] (11.4,0.5)--(0,0.5);
	\node at (-0.5,0.5) {$\rho_{\varphi 0}$};
	\draw [dashed] (11.4,-1.1)--(0,-1.1);
	\node at (-0.5,-1.1) {$\rho_\textrm{r0}$};
	\draw [dashed] (2.55,4.3)--(0,4.3);
	\node at (-0.5,4.3) {$\rho_\textrm{re}$};
	\node at (5,2)  {$\rho_\varphi\propto a^{-6}$};
	\node at (6,3.0) {$\rho_\textrm{r}\propto a^{-4}$};
	\node at (9.5,1.1) {$\rho_\textrm{m}\propto a^{-3}$};
	\node at (7,0.8)  {$\rho_\varphi\propto a^{0}$};
	\end{tikzpicture}
	\caption{A schematic plot of the energy densities of the inflaton field, radiation and matter after the end of inflation, respectively denoted by $\rho_\varphi$, $\rho_\textrm{r}$ and $\rho_\textrm{m}$. The evolution of the matter energy density before the matter-radiation equality is not relevant for our purpose, so it is not shown to avoid making the plot busy.   We used the same notations for scale factors as in figure \ref{fig_aH}. Gradients of the lines were exaggerated for illustrative purpose. One should take the evolution of $\rho_\varphi$ and $\rho_\textrm{r}$ during reheating with a grain of salt, since they could be altered by the interaction between the inflaton field and radiation.} \label{fig_rho}
\end{figure}

Figure \ref{fig_rho} sketches the evolution of energy densities of inflaton, radiation and matter, denoted by $\rho_\varphi$, $\rho_\textrm{r}$ and $\rho_\textrm{m}$, ever since the inflation was over.
During the reheating, the energy density of the inflaton field more rapidly decreases than that of radiation and the reheating terminates when these energy densities become equal. After the reheating, the inflaton field and radiation are decoupled from each other and stream freely. While the radiation energy density scales as $a^{-3}$ to the present time, the inflaton energy density scales as $a^{-6}$ in the kinetic energy dominated phase after which it remains nearly constant and finally catches up the radiation-matter energy density to drive the late-time accelerated expansion. 

The evolution of the universe after reheating was systematically investigated in the literature, see e.g. \cite{Akrami:2017cir}, by tracking the dynamics of the inflaton field \textit{forward} in time with the initial conditions at early times set by hand. 

However, we can in principle go \textit{backward} in time to determine the energy densities of the inflaton field and radiation at any moment after reheating by using the final conditions at the present time determined by observations, since we know the energy densities of dark energy, matter and radiation today from observations. Apart from the final conditions set by observational parameters, more importantly, this will allow us to determine the scale factor $a_\textrm{re}$ at the end of reheating without knowledge of reheating mechanisms by solving
\begin{equation}\label{rho_phi-r}
	\rho_\textrm{re}=\rho_\varphi(a_\textrm{re})=\rho_\textrm{r}(a_\textrm{re})=\rho_\textrm{r0}\left(\frac{a_\textrm{re}}{a_0}\right)^{-4},
\end{equation}
where $\rho_\textrm{r0}\equiv \rho_\textrm{r}(a_0)$.
Once $a_\textrm{re}$ is determined, we can compute $\rho_\textrm{re}$ from eq. \eqref{rho_phi-r}, which means that point Q in figure \ref{fig_aH} is fixed, or equivalently the equation of state parameter $w_\textrm{re}$ during reheating is determined. Explicitly, we have
\begin{equation}\label{wre}
	w_\textrm{re}=\frac{1}{3}\log_{\frac{a_\textrm{re}}{a_\textrm{end}}}{\frac{\rho_\textrm{end}}{\rho_\textrm{re}}}-1.
\end{equation}
Knowing $a_\textrm{re}$ greatly simplifies the process to compute the reheating temperature and the number of e-folds during reheating. Indeed, the reheating temperature directly follows from eq. \eqref{Tre-T0} that
\begin{equation}\label{Tre-qi}
	T_\textrm{re}=\left(\frac{43}{11g_{s,\textrm{re}}}\right)^{1/3}\frac{a_0}{a_\textrm{re}}T_0,
\end{equation} 
 and the number of e-folds during reheating is
\begin{equation}\label{Nre-qi}
	N_\textrm{re}=\ln(\frac{a_\textrm{re}H_k}{k_*})-N_k.
\end{equation}

In order to find $a_\textrm{re}$ we should address how to express the inflaton energy density as a function of the scale factor. It can be done by solving the following simultaneous equations 
\begin{equation}\label{KG_N}
	\frac{d}{dN}\left[H(N)\varphi'(N)\right]+3H(N)\varphi'(N)+\frac{V_\varphi[\varphi(N)]}{H(N)}=0,
\end{equation}
\begin{equation}
	H(N)=\sqrt{\frac{\rho_\varphi(N)+\rho_\textrm{r}(N)+\rho_\textrm{m}(N)}{3}},\label{H_N}
\end{equation}
\begin{equation}
\rho_\varphi(N)=\frac{H^2(N)\varphi'^2(N)}{2}+V[\varphi(N)],\label{rho-phi_N}
\end{equation}
\begin{equation}
\rho_\textrm{r}(N)=\rho_\textrm{r0}\,e^{-4N},\label{rho-r_N}
\end{equation}
\begin{equation}
\rho_\textrm{m}(N)=\rho_\textrm{m0}\,e^{-3N}.\label{rho-m_N}
\end{equation}
Here we used a prime to indicate the derivative with respect to the number of e-folds $N\equiv \ln \frac{a}{a_0}$. The energy densities of radiation and matter at present are 
\begin{equation}
\rho_\textrm{r0}=3H_0^2\Omega_\textrm{r},
\end{equation}
\begin{equation}
\rho_\textrm{m0}=3H_0^2\Omega_\textrm{m},
\end{equation}
where $\Omega_\textrm{r}$ and $\Omega_\textrm{m}$ denote the density parameters today for radiation and matter, respectively.
We rewrite eqs. \eqref{BC-1} and \eqref{BC-2} in terms of the number of e-folds 
\begin{equation}
	\rho_{\varphi 0}=3H_0^2\Omega_\Lambda,
\end{equation}
\begin{equation}
	\frac{\frac{H_0^2\varphi'^2(0)}{2}-V(\varphi_0)}{\frac{H_0^2\varphi'^2(0)}{2}+V(\varphi_0)}=w_0,
\end{equation}
which reduce to the final conditions
\begin{equation}\label{phi0}
	V(\varphi_0)=\frac{3\Omega_\Lambda H_0^2(1-w_0)}{2},
\end{equation}
\begin{equation}\label{dotphi0}
	\varphi'(0)=\sqrt{3\Omega_\Lambda(1 + w_0)}.
\end{equation}
It now seems that we are all set for solving eqs. \eqref{KG_N}--\eqref{rho-m_N}. However, in practice, the value of the inflaton field today $\varphi_0$ can not be uniquely determined, since eq. \eqref{phi0} admits a large range of $\varphi_0$ when the observational uncertainties are taken into account, due to the asymptotic behavior of potentials. Hence, $\varphi_0$ will be taken as a free parameter for our analysis, though there are model-dependent constraints on it.

After reheating, the kinetic energy of the inflaton field dominates over its potential energy so that the potential gradient term in eq. \eqref{KG_N} becomes negligible compared to the Hubble friction term. Accordingly, the inflaton field slows down and gets frozen. When the inflaton field starts to feel the potential gradient, it unfreezes and finally moves with a constant velocity ever since the last two terms in eq. \eqref{KG_N} become equal. The point is that the moment when the inflaton field unfreezes is not known a priori: it can come earlier or later than the present time depending on $\varphi_0$. Namely, there are two cases: (i) the inflaton field remains frozen until today; (ii) the inflaton field unfreezes and rolls at the present time.
 We focus here on the first case for at least the following two reasons:
  \begin{itemize}
 	\item The cosmic observations \cite{planckR} indicate that the equation of state of dark energy today is almost equal to $-1$, which can be better explained when the inflaton field remains almost frozen today.
 	\item Given above, one may ask whether the equation of state of dark energy can be \textit{arbitrarily} close to $-1$ or there exists any bound on it. 
 \end{itemize}

Below we will show that the consistency with reheating imposes very tight bounds on the equation of state of dark energy today.
 
 In the case when the inflaton field freezes until today, one can neglect the gradient of the potential all the way to the present time.\footnote{Very much similar behavior appears in the early universe, which is dubbed the ultra slow-roll inflation \cite{Martin:2012pe,Dimopoulos:2017ged}.} Then we can simplify the equation \eqref{KG_N} by dropping the last term,
\begin{equation}\label{KG_N-V0}
\frac{d}{dN}\left[H(N)\varphi'(N)\right]+3H(N)\varphi'(N)\simeq 0.
\end{equation}
With the final condition \eqref{dotphi0}, we can solve eq. \eqref{KG_N-V0} to obtain
\begin{equation}
H(N)\varphi'(N)=H_0\sqrt{3(1 + w_0)\Omega_\Lambda}\,e^{-3N},
\end{equation} 
which can be used in eq. \eqref{rho-phi_N} to yield the inflaton energy density at the end of reheating
\begin{equation}\label{rhophi_re}
\rho_\varphi(a_\textrm{re})=\frac{3(1+w_0)\Omega_\Lambda}{2}H_0^2a_\textrm{re}^{-6}.
\end{equation}
Here we used that the kinetic energy of the inflaton field dominates the potential energy at the end of reheating.
After the substitution of eq. \eqref{rhophi_re} into eq. \eqref{rho_phi-r}, we find
\begin{equation}\label{are-app}
a_\textrm{re}=\sqrt{\frac{\Omega_\Lambda}{\Omega_\textrm{r}}\left(\frac{1+w_0}{1-w_0}\right)}.
\end{equation}
It then follows from eq. \eqref{Tre-qi} that the reheating temperature is
\begin{equation}\label{Tre-app}
T_\textrm{re}=\left(\frac{43}{11g_{s,\textrm{re}}}\right)^{1/3}\sqrt{\frac{\Omega_\textrm{r}}{\Omega_\Lambda}\left(\frac{1-w_0}{1+w_0}\right)}\,T_0.
\end{equation}

Eq.  \eqref{Tre-app} shows that the reheating temperature in quintessential  inflation models can be determined merely from late-time observational parameters, namely $\Omega_\Lambda$, $\Omega_\textrm{r}$, $w_0$ and $T_0$.
  Conversely, if we know the reheating temperature somehow, by inverting eq. \eqref{Tre-app} we can determine the equation of state parameter of the inflaton field today, i.e.
 \begin{equation}\label{w0-app}
 1+w_0\simeq\left(\frac{43}{11g_{s,\textrm{re}}}\right)^{2/3}\,\frac{2\Omega_\textrm{r}}{\Omega_\Lambda}\left(\frac{T_0}{T_\textrm{re}}\right)^2,
 \end{equation}	
where we have used  $1-w_0\simeq 2$.
 	 For instance, the BBN constraint on the reheating temperature $T_\textrm{re}\gtrsim 1\textrm{MeV}$ \cite{Giudice:2000ex} gives the universal upper bound 
\begin{equation}\label{w0_UB}
w_0\lesssim -1+10^{-24}
\end{equation}
for  $\Omega_\Lambda=0.685$, $\Omega_\textrm{r}=9.0\times 10^{-5}$ and $g_{s,\textrm{re}}=106.75$ \cite{planckR}.
This quantitatively tells us that quintessential inflation models with successful reheating could be indistinguishable from the cosmological constant unless the inflaton field unfreezes before the present time. 
The lower bound of $w_0$ can be estimated as
\begin{equation}\label{w0_LB}
w_0\gtrsim -1+10^{-60},
\end{equation}
by using that $T_\textrm{re}\lesssim 10^{15}\textrm{GeV}$. The extremely closeness of $w_0$ to $-1$ has its root in a large temperature hierarchy between the end of reheating and the present time.
One may obtain a more stringent lower bound of $w_0$ by using $w_\textrm{re}\leq 1$, but in a model-dependent way due to $w_\textrm{re}$ being a model-dependent quantity (see eq. \eqref{wre}). 

Another interesting implication of eq. \eqref{Tre-app} is that it can provide a link between inflationary predictions and late-time observational parameters. Indeed, combining eqs. \eqref{Tre-QI} and \eqref{Tre-app}, one can find
\begin{equation}
\mathcal{N}(\mathcal{P}_i,\mathcal{O}_i)-\frac{1}{12}\ln\left(\frac{40V_\textrm{end}}{\pi^2g_{*,\textrm{re}}T_0^4}\right)-\frac{1}{9}\left(\frac{11g_{s,\textrm{re}}}{43}\right)=\frac{1}{6}\ln\left[\frac{\Omega_\Lambda}{\Omega_\textrm{r}}\left(\frac{1+w_0}{1-w_0}\right)\right] .\label{ns-w0}
\end{equation}
It is manifest that the left-hand side of eq. \eqref{ns-w0} contains the inflationary predictions such as the scalar spectral index and the tensor-to-scalar ratio, while the right-hand side includes only late-time observational parameters.

We emphasize that eq.  \eqref{Tre-app} has been derived without assuming a specific reheating mechanism and inflation model, so it is the universal relation that can apply to any quintessential inflation model, provided that the inflaton field remains frozen today. This universality will be numerically checked for various models in section \ref{sec-examples}. 

\section{Examples}\label{sec-examples}
In this section we consider several quintessential inflation models  widely studied in the literature \cite{Akrami:2017cir,Alho:2023pkl,Sarkar:2023cpd,Salo:2021piz,Dimopoulos:2017tud,Brissenden:2023yko,Dimopoulos:2017zvq,Garcia-Garcia:2018hlc,Garcia-Garcia:2019ees} to demonstrate the existence of physically viable regime in which 
the gradient of the inflaton potential could be neglected until today ever since the reheating was over. We also confirm numerically the analytic relation \eqref{Tre-app} between the reheating temperature and the late-time observational parameters. For all models to be discussed in this section, we will choose the current value of the inflaton field in such a way that the potential gradient is negligible compared to the Hubble friction and the elongation of the inflaton field obeys the constraint $|\varphi_0-\varphi_\textrm{end}|\lesssim 43$ \cite{Dimopoulos:2017zvq}.  

\subsection{$\alpha$-attractor model with linear potential}
The simplest potential of $\alpha$-attractor quintessential inflation models is the linear potential \cite{Akrami:2017cir} which in terms of the canonically normalized field $\varphi$ is written as
\begin{equation}\label{V_linear}
	V(\varphi)=\gamma\sqrt{6\alpha}\left[\tanh(\frac{\varphi}{\sqrt{6\alpha}})+1\right]+\Lambda,
\end{equation}
where $\gamma$ and $\alpha$ are model parameters and $\Lambda$ the cosmological constant observationally constrained by $\Lambda=\frac{3H_0^2\Omega_\Lambda(1-w_0)}{2}$. It has two nearly flat plateaus at $\varphi\gg \sqrt{6\alpha}$ and $\varphi\ll-\sqrt{6\alpha}$ corresponding to inflation and the late-time accelerated expansion.

\subsubsection{Inflationary predictions}
Let us first derive the predictions of the model on inflationary parameters. The typical scale of the potential \eqref{V_linear} during inflation is $V(\varphi)\sim 2\gamma\sqrt{6\alpha}$. Then the upper bound of the tensor-to-scalar ratio $r<0.032$ \cite{Tristram:2021tvh} and the Hubble parameter at the horizon crossing \eqref{Hk-phik} confine model parameters by
\begin{equation}
	2\gamma\sqrt{6\alpha}\lesssim 10^{-9}.
\end{equation}
 Neglecting the cosmological constant in the potential \eqref{V_linear} during inflation and using it in eqs. \eqref{Hk-phik} and \eqref{phi_end}, we get
\begin{equation}
	\varphi_k=\sqrt{\frac{3\alpha}{2}}\ln(\frac{-1+\sqrt{1+\frac{16\pi^2A_s}{\gamma\sqrt{6\alpha^3}}}}{2}),
\end{equation}
\begin{equation}
\varphi_\textrm{end}=\sqrt{\frac{3\alpha}{2}}\ln(\sqrt{\frac{1}{3\alpha}}-1),\label{phi_end_linear}
\end{equation}
which gives the number of e-folds during inflation 
\begin{align}
	N_k\simeq \sqrt{\frac{3\pi^2\sqrt{6\alpha}A_s}{2\gamma}}.\label{Nk_linear}
\end{align}
Note from eq. \eqref{phi_end_linear} that a graceful exit from inflation is possible only for $\alpha<1/3$. The scalar spectral index is 
\begin{equation}\label{ns_linear}
	n_s=1-6\epsilon(\varphi_k)+2\eta(\varphi_k)\simeq 1-\sqrt{\frac{8\gamma}{3\pi^2\sqrt{6\alpha}A_s}}.
\end{equation}
Using eq. \eqref{ns_linear}, we can express $\gamma$ as
\begin{equation}
	\gamma=\frac{3}{4}\sqrt{\frac{3}{2}}A_s(1-n_s)^2\pi^2\sqrt{\alpha}.
\end{equation}
Then the CMB constraint on the scalar spectral index $0.9607<n_s<0.9691$ \cite{Planck:2018jri} at $1\sigma$ confidence level puts a stringent constraint on $\gamma$ for a given $\alpha$
\begin{equation}
	1.8\times 10^{-11}\sqrt{\alpha}<\gamma<2.94\times 10^{-11} \sqrt{\alpha}.
\end{equation}

\subsubsection{Inflaton dynamics after reheating}\label{numerics-linear}
Now we turn our attention to the late-time dynamics of the inflaton field. It is not difficult to see from eq. \eqref{V_linear} that the potential is dominated by the cosmological constant when 
\begin{equation}
	\varphi<\varphi_\Lambda\equiv\sqrt{6\alpha}\tanh^{-1}\left(\frac{\Lambda}{\sqrt{6\alpha}\gamma}-1\right).\label{phi_Lambda}
\end{equation}
Also, the gradient of potential \eqref{V_linear} today is negligible for
\begin{equation}
	\varphi_0<-\sqrt{6\alpha} \cosh^{-1}\left(\sqrt{\frac{\gamma}{3H_0^2\sqrt{3\Omega_\Lambda(1+w_0)}}}\right).\label{phi_grad}
\end{equation}

\begin{figure}[!htb]
	\centering
	\begin{subfigure}{0.45\textwidth}
\includegraphics[scale=0.38]{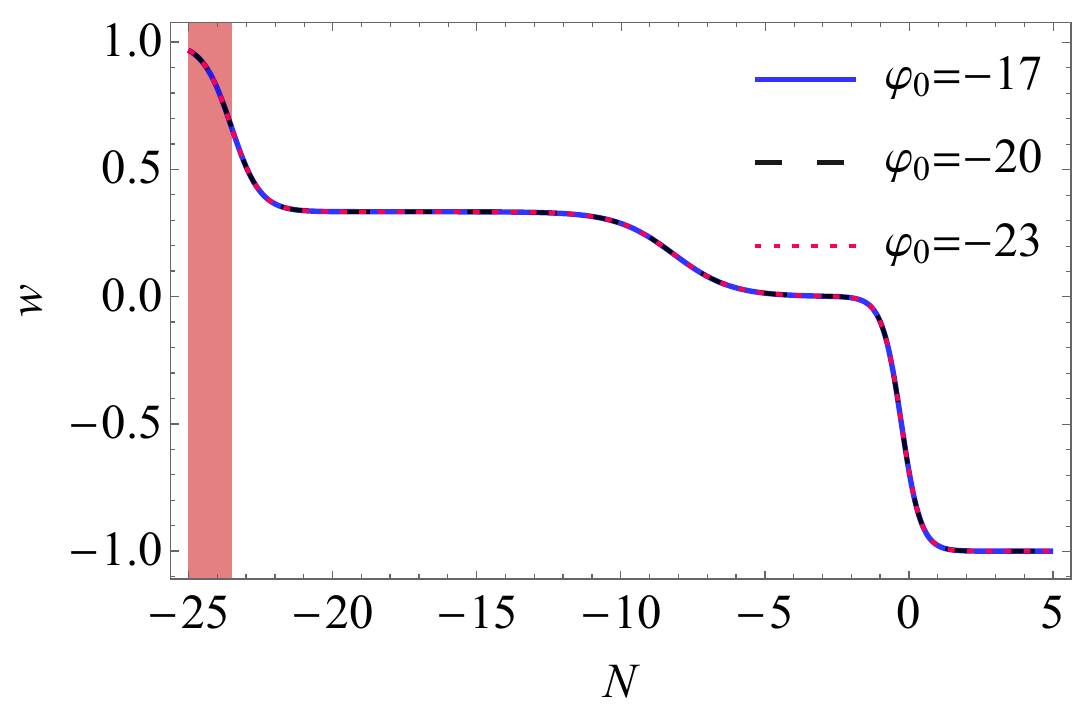}
\subcaption{}\label{fig_w-N-1}
	\end{subfigure}
\begin{subfigure}{0.45\textwidth}
	\includegraphics[scale=0.38]{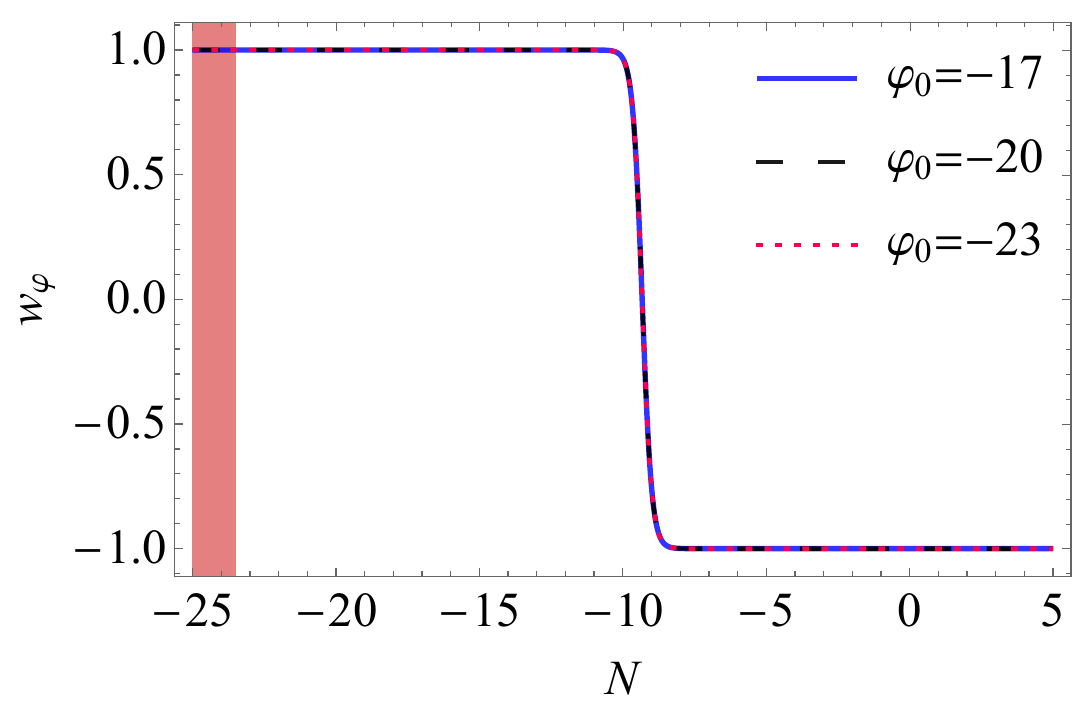}
	\subcaption{}\label{fig_wphi-N-1}
\end{subfigure}
\begin{subfigure}{0.45\textwidth}
	\includegraphics[scale=0.38]{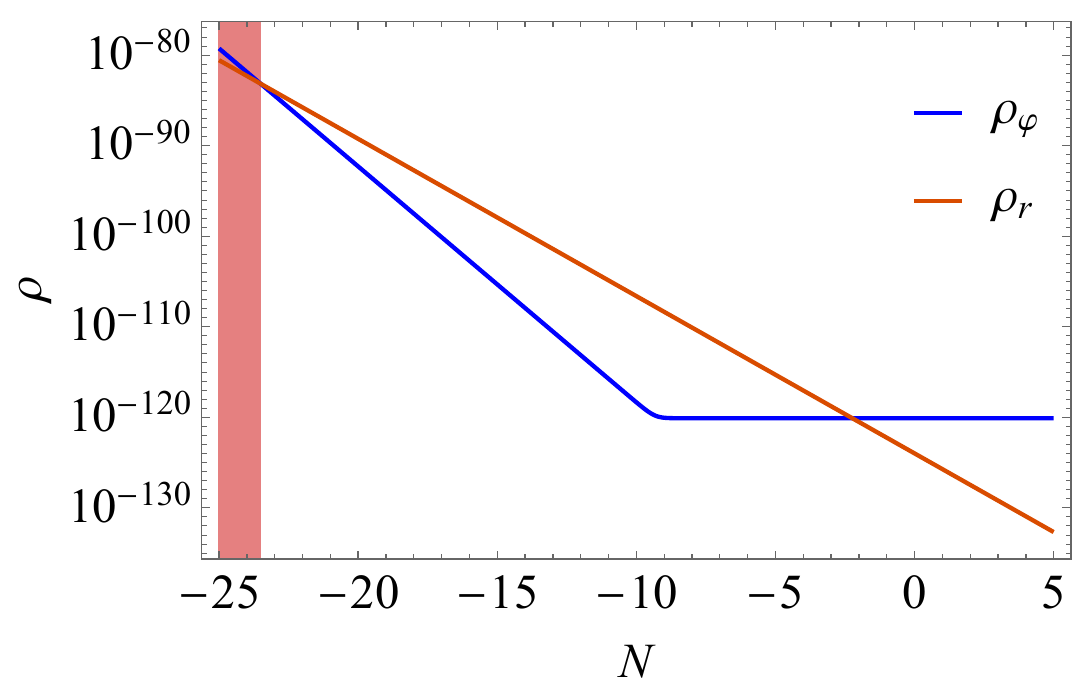}
	\subcaption{}\label{fig_rho-N-1}
\end{subfigure}
\begin{subfigure}{0.45\textwidth}
	\includegraphics[scale=0.38]{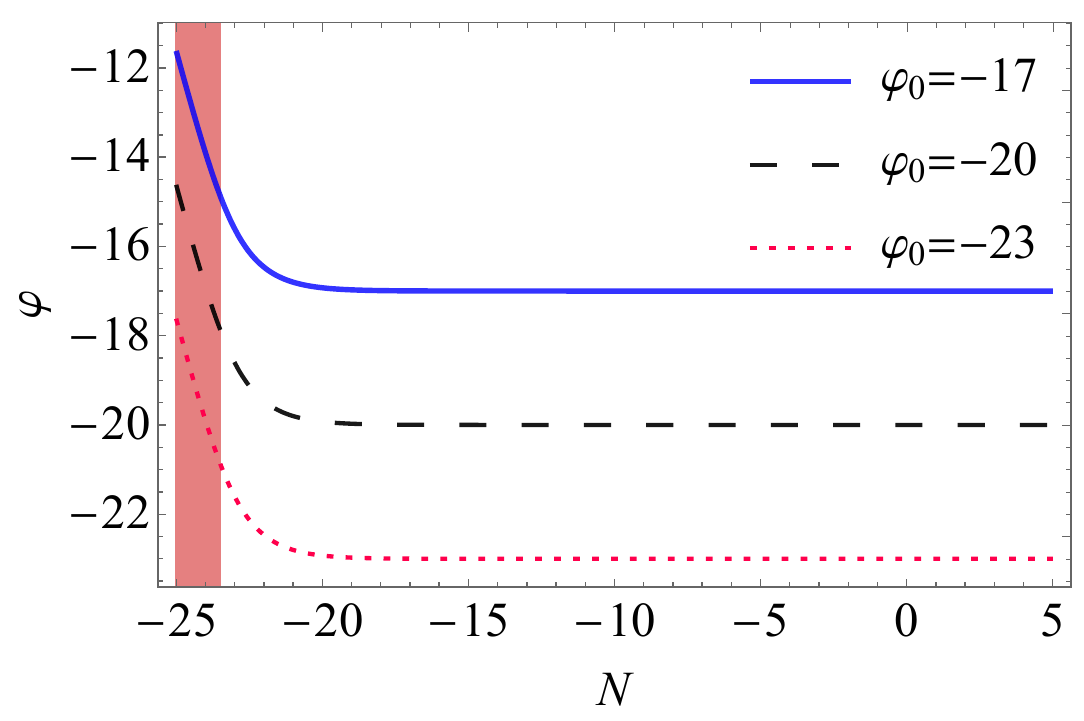}
	\subcaption{}\label{fig_phi-N-1}
\end{subfigure}
\caption{Dynamics of the universe after reheating in the $\alpha$-attractor quintessential inflation models with the linear potential \eqref{V_linear}. We set the model parameters to $\alpha=0.002$ and $\gamma=8.5\times 10^{-13}$ and used the cosmological parameters today $\Omega_\Lambda=0.685$, $w_0=-1+10^{-24}$, $\Omega_\textrm{r}=9.0\times 10^{-5}$ and $H_0=5.9\times 10^{-61}$. The reheating period was indicated by the shaded region where one should not trust the numeric results, since the dynamics of the inflaton field and radiation could be affected by particle production mechanisms. (a) The evolution of effective equation of state parameter of the universe. (b) The evolution of the equation of state parameter of the inflaton field. (c) The evolution of the energy densities of inflaton field and radiation. (d) The evolution of the inflaton field for various final conditions.}\label{fig:linear-1}
	\end{figure}

\begin{figure}[!htb]
	\centering
	\begin{subfigure}{0.45\textwidth}
		\includegraphics[scale=0.38]{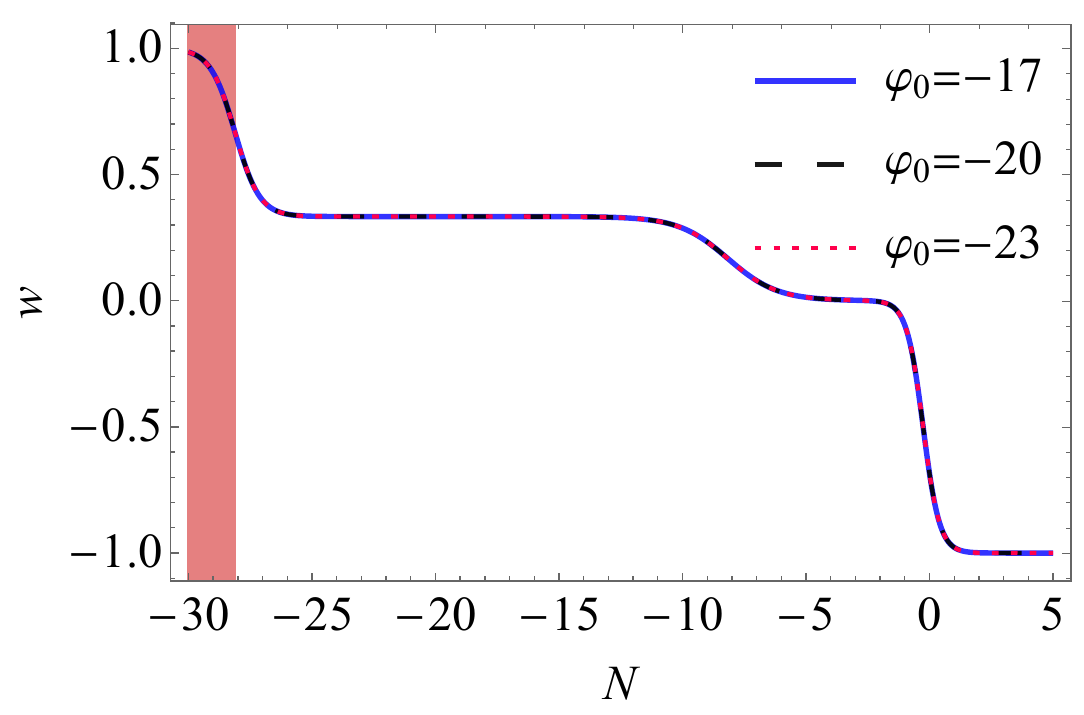}
		\subcaption{}\label{fig_w-N-2}
	\end{subfigure}
	\begin{subfigure}{0.45\textwidth}
		\includegraphics[scale=0.38]{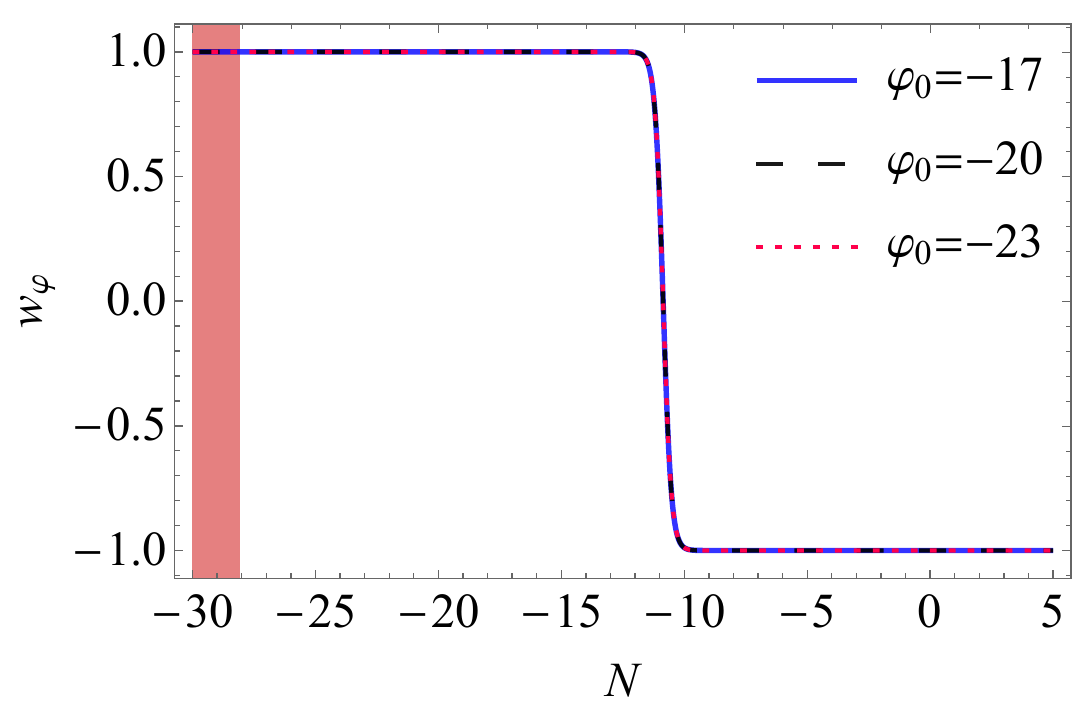}
		\subcaption{}\label{fig_wphi-N-2}
	\end{subfigure}
	\begin{subfigure}{0.45\textwidth}
		\includegraphics[scale=0.38]{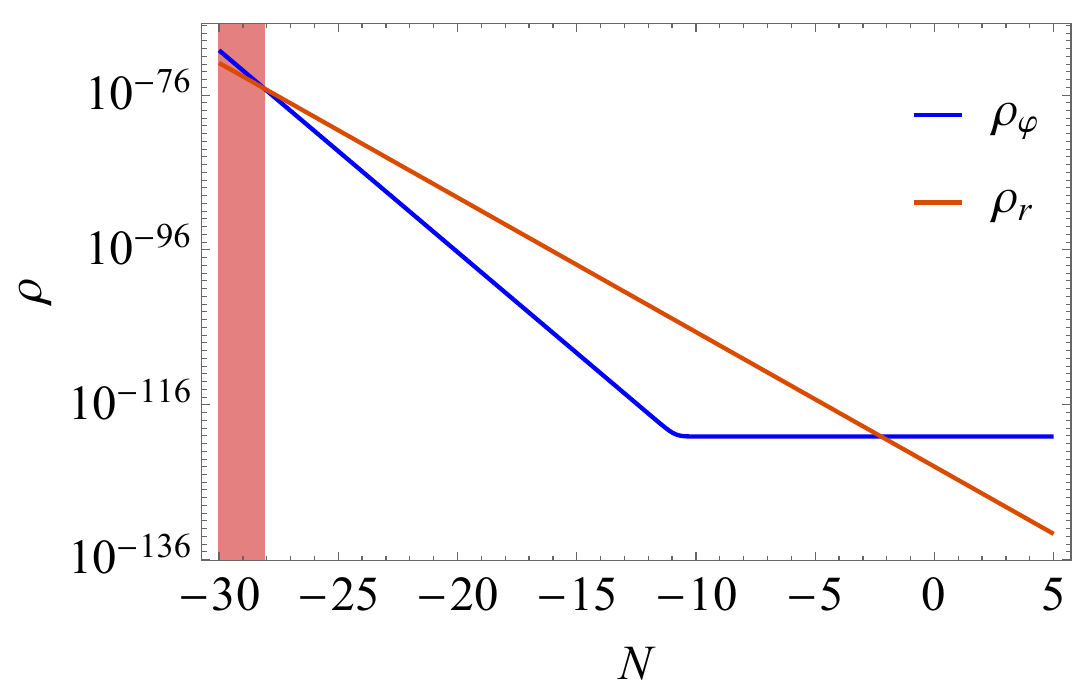}
		\subcaption{}\label{fig_rho-N-2}
	\end{subfigure}
	\begin{subfigure}{0.45\textwidth}
		\includegraphics[scale=0.38]{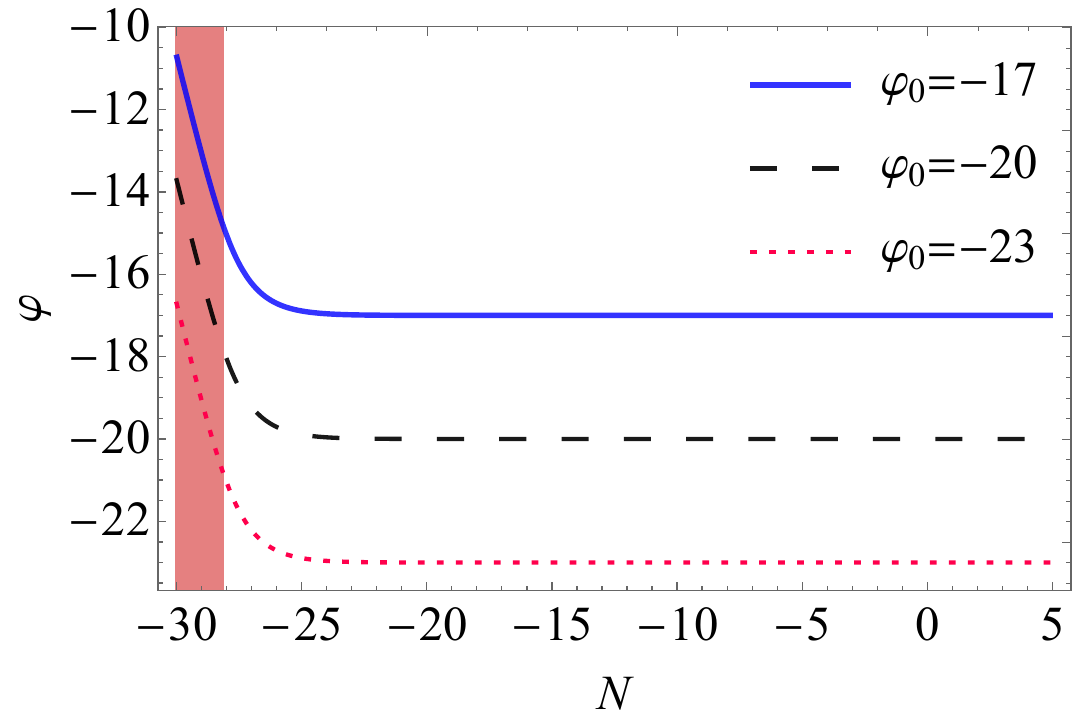}
		\subcaption{}\label{fig_phi-N-2}
	\end{subfigure}
	\caption{The same as figure \ref{fig:linear-1}, but with $w_0=-1+10^{-28}$.}\label{fig:linear-2}
\end{figure}

We choose the current value of the inflaton field $\varphi_0$ such that the final condition \eqref{phi0} together with the constraints \eqref{phi_Lambda} and \eqref{phi_grad} are satisfied.  With the potential \eqref{V_linear} and the various final conditions on $\varphi_0$ and $\varphi'(0)$, we numerically solve the simultaneous equations \eqref{KG_N}--\eqref{rho-m_N}. The model parameters are set to $\alpha=0.002$ and $\gamma=8.5\times 10^{-13}$. The results are shown in figures \ref{fig:linear-1} and  \ref{fig:linear-2}. These figures exhibit important features of the model.

\begin{itemize}
	\item As shown in figures \ref{fig_w-N-1}, \ref{fig_wphi-N-1}, \ref{fig_w-N-2}  and \ref{fig_wphi-N-2}, the model can describe  correctly the evolution of the universe after reheating when the equation of state parameter of the inflaton field is extremely close to $-1$ at the present time.\footnote{In the context of Chevallier-Polarski-Linder (CPL) parameterization \cite{CHEVALLIER_2001,0208512} $w_\varphi(z)=w_0+w_az/(1+z)$ with $1+z\equiv e^{-N}$ valid for $-1\lesssim N\lesssim 0$,  figures \ref{fig_wphi-N-1} and  \ref{fig_wphi-N-2} indicate  $w_a\simeq 0$ so that the model under consideration would be indistinguishable from the cosmological constant.} The reason is the following. The greater the deviation of $w_0$ from $-1$, the greater the current value of the velocity of the inflaton field. This implies that the moment after which the potential energy dominates the inflaton energy density is too close to the present time so that the model predicts too short period of RD leading to the incorrect evolution of the universe, unless $w_0$ is very close to $-1$.
	
	\item The evolution of the universe after reheating is less sensitive to the current value of the inflaton field. Though the value of the inflaton field at given moment depends on the final condition (see figures \ref{fig_phi-N-1} and \ref{fig_phi-N-2}), the equation of state parameters of the universe and the inflaton field are not affected as can be seen from figures \ref{fig_w-N-1}, \ref{fig_wphi-N-1}, \ref{fig_w-N-2} and \ref{fig_wphi-N-2}. This is partly due to fact that at the present time the velocity of the inflaton field does not depend on the  value of the inflaton field but $w_0$ (see eq. \eqref{dotphi0}) and partly because of the negligible potential gradient.	
	
	\item The end of reheating at which the energy densities of the inflaton field and radiation become equal, is sensitive to the equation of state parameter $w_0$ of dark energy today. Figures \ref{fig_rho-N-1} and  \ref{fig_rho-N-2} show that the reheating ends earlier for $w_0$ closer to $-1$. 
	
\end{itemize}

By using numerical solutions to eqs. \eqref{KG_N}--\eqref{rho-m_N} for a given model parameters, we can determine the scale factor $a_\textrm{re}$ at the end of reheating at which we have $\rho_\varphi(a_\textrm{re})=\rho_\textrm{r}(a_\textrm{re})$ (cf. figures \ref{fig_rho-N-1} and  \ref{fig_rho-N-2}). In particular we obtain for the model parameters $\alpha=0.002$ and $\gamma=8.5\times 10^{-13}$
\begin{align}
	a_\textrm{re}&=6.2\times 10^{-11} \quad \mbox{for $w_0=-1+10^{-24}$},\label{are-linear-1}\\
	a_\textrm{re}&=6.2\times 10^{-13} \quad \mbox{for $w_0=-1+10^{-28}$}.\label{are-linear-2}
\end{align}
Also, by using eqs. \eqref{are-linear-1} and \eqref{are-linear-2} in eqs.  \eqref{Tre-qi}, we get the reheating temperature\footnote{Note here that the reheating temperature is written in the natural units $c=\hbar=1$ for easy comparison to the BBN constraint.}
\begin{align}
T_\textrm{re}&=1.3\textrm{MeV} \quad \mbox{for $w_0=-1+10^{-24}$},\label{Tre-linear-1}\\
T_\textrm{re}&=0.13\textrm{GeV} \quad \mbox{for $w_0=-1+10^{-28}$}.\label{Tre-linear-2}
\end{align}

One can easily check that the numerical results \eqref{are-linear-1}--\eqref{Tre-linear-2} are in good agreement with the  analytic relations \eqref{are-app} and \eqref{Tre-app}. 

\subsection{$\alpha$-attractor model with exponential potential}

Let us consider a quintessential  $\alpha$-attractor inflation model with the exponential potential \cite{Dimopoulos:2017zvq,Akrami:2017cir} 
\begin{equation}\label{potential}
	V(\varphi)=M^2e^{\gamma\left[\tanh(\frac{\varphi}{\sqrt{6\alpha}})-1\right]}+V_0,
\end{equation} 
where $M$, $\gamma$, $\alpha$ and $V_0$ are free parameters. Two specific subclasses of this model are Exp-model I and 	Exp-model II \cite{Akrami:2017cir} for which we choose $V_0=0$ and $V_0=-M^2e^{-2\gamma}$, respectively. In Exp-model II, the equation of state parameter of the inflaton field today $w_0\simeq 0.5(1+w_\infty)=-1+\frac{1}{9\alpha}$ \cite{Linder:2023}, which implies $\alpha>27.8$ to agree with the observation  $w_0=-1.028\pm 0.032$. However, such a large value of $\alpha$ results in the tensor-to-scalar ratio $r=\frac{12\alpha}{N_k^2}\sim 0.07$ 
which lies the outside of the region $r<0.032$ allowed by observations \cite{Tristram:2021tvh}.
So we here focus on Exp-model I which has the potential 
\begin{equation}
	V(\varphi)=M^2e^{\gamma\left[\tanh(\frac{\varphi}{\sqrt{6\alpha}})-1\right]}.\label{V_exp}
\end{equation}
As in the case of linear potential, it drives the inflation for $\varphi\gg {\sqrt{6\alpha}}$ and the late-time accelerated expansion for $\varphi\ll -{\sqrt{6\alpha}}$.

\subsubsection{Inflationary predictions}

We can approximate the potential \eqref{V_exp} during inflation as
\begin{equation}
V(\varphi)=M^2\exp\left(-2\gamma e^{-\sqrt{\frac{2}{3\alpha}}\varphi}\right),\label{Vinf_exp}
\end{equation}
from which we read that $M$ determines the inflation energy scale so that it obeys
\begin{equation}
	M^2\lesssim 10^{-10}.
\end{equation}
Here we used $r<0.032$ with eq. \eqref{Hk-phik}.
Using eqs. \eqref{Hk-phik} and \eqref{phi_end} with the potential \eqref{Vinf_exp}, we can write $\varphi_k$ and $\varphi_\textrm{end}$ in terms of model parameters
\begin{equation}
	\varphi_k\simeq \sqrt{\frac{3\alpha}{2}}\ln(\frac{4\pi\gamma\sqrt{2A_s}}{M\sqrt{\alpha}}),\label{phik_exp}
\end{equation}
\begin{equation}
	\varphi_\textrm{end}\simeq\sqrt{\frac{3\alpha}{2}}\ln(\frac{2\gamma}{\sqrt{3\alpha}}).\label{phiend_exp}
\end{equation}
Accordingly, the number of e-folds during inflation is given by
\begin{equation}
	N_k\simeq \frac{3\pi\sqrt{2\alpha A_s}}{M}.
\end{equation}
We substitute eqs. \eqref{phik_exp} and \eqref{phiend_exp} into eq. \eqref{slowroll parameter} to obtain the scalar-spectral index as
\begin{equation}\label{ns_exponential}
n_s=1-6\epsilon(\varphi_k)+2\eta(\varphi_k)\simeq 1-\frac{M^2+4\pi M\sqrt{\frac{2A_s}{\alpha}}}{12\pi^2A_s}.
\end{equation}
This implies that parameters $\alpha$ and $M$ are no longer independent but constrained by the Planck result $0.9607<n_s<0.9691$ \cite{Planck:2018jri} at $1\sigma$ confidence level.

\subsubsection{Inflaton dynamics after reheating}\label{numerics-exp}
\begin{figure}[!htb]
	\centering
	\begin{subfigure}{0.45\textwidth}
		\includegraphics[scale=0.4]{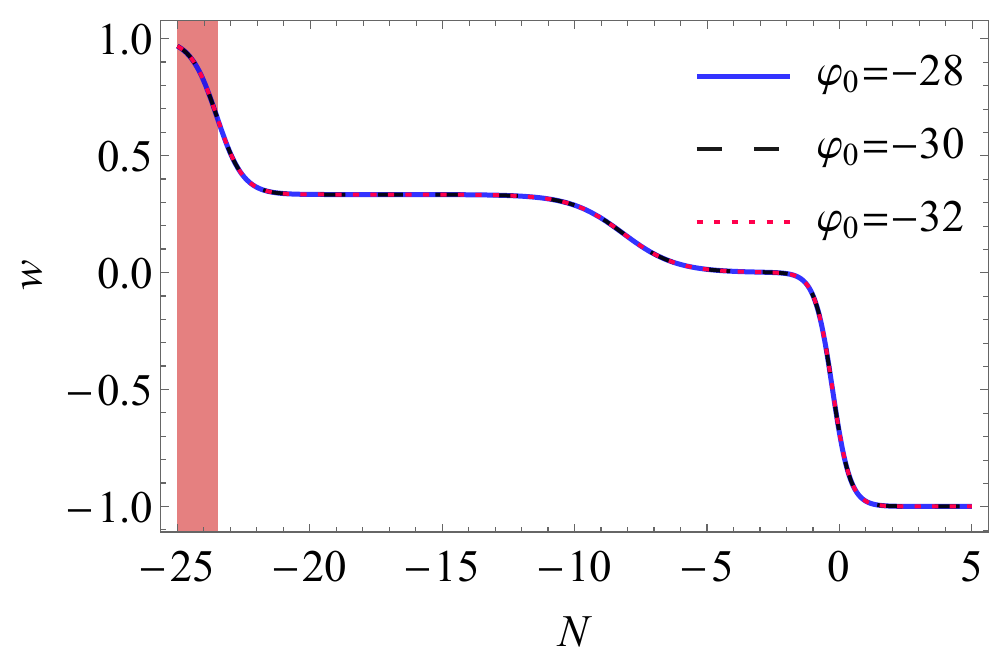}
		\subcaption{}\label{fig_exp_w-N-1}
	\end{subfigure}
	\begin{subfigure}{0.45\textwidth}
		\includegraphics[scale=0.4]{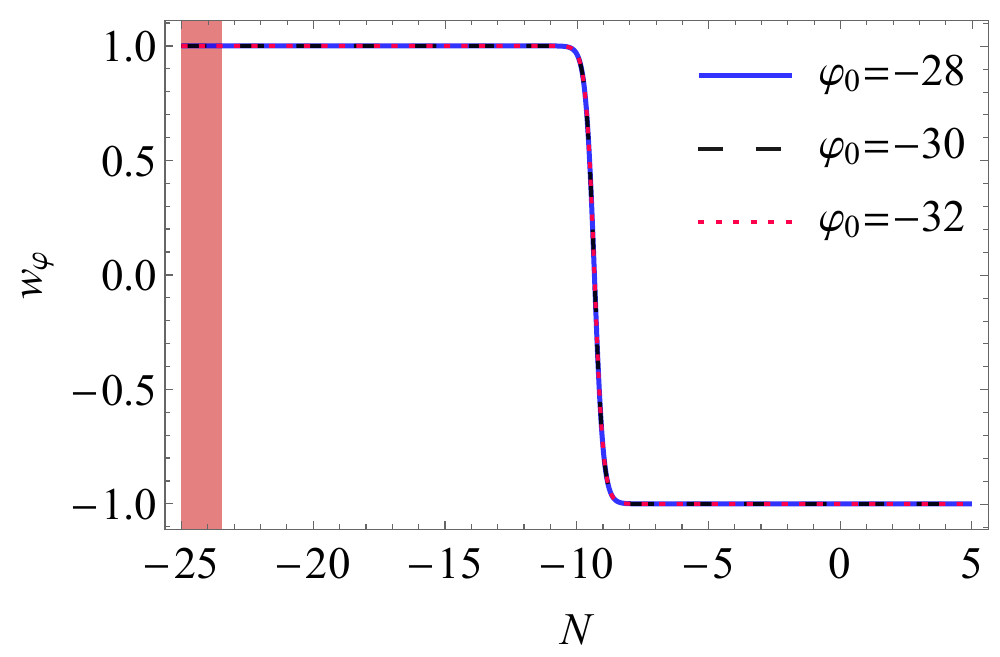}
		\subcaption{}\label{fig_exp_wphi-N-1}
	\end{subfigure}
	\begin{subfigure}{0.45\textwidth}
		\includegraphics[scale=0.4]{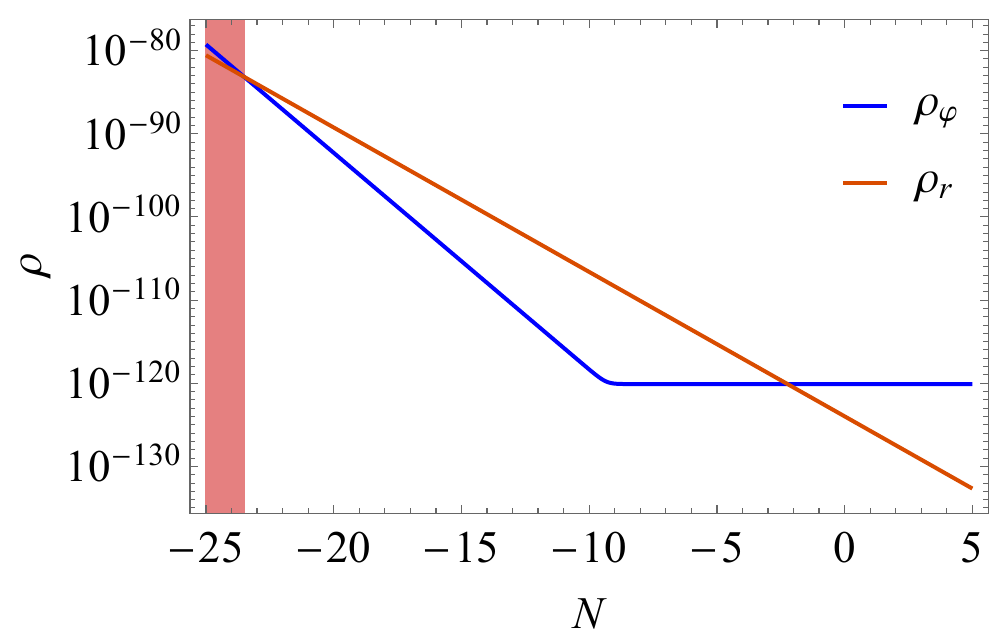}
		\subcaption{}\label{fig_exp_rho-N-1}
	\end{subfigure}
	\begin{subfigure}{0.45\textwidth}
		\includegraphics[scale=0.4]{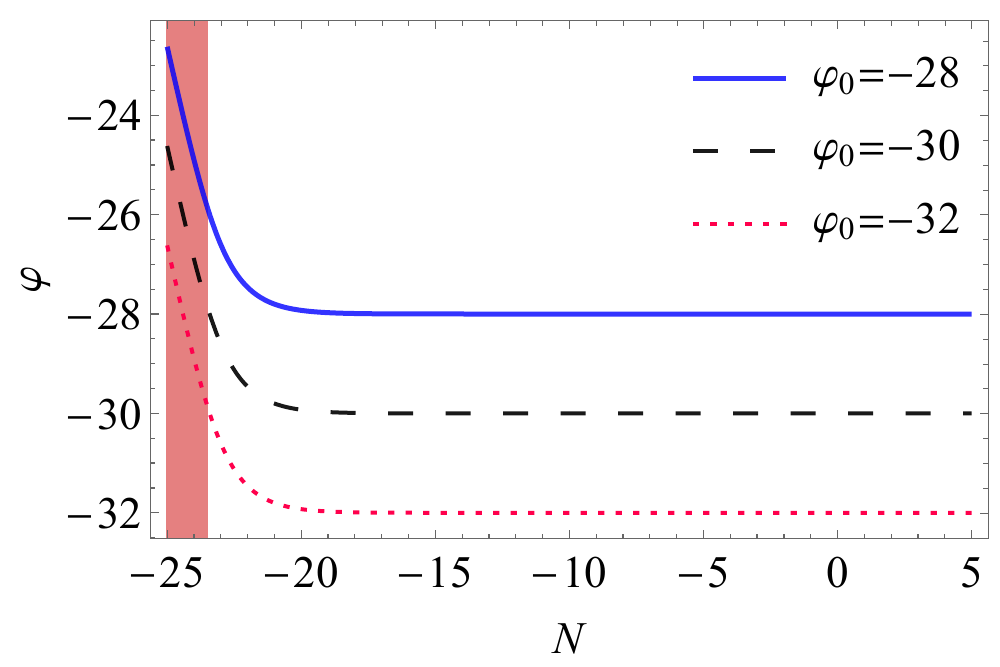}
		\subcaption{}\label{fig_exp_phi-N-1}
	\end{subfigure}
	\caption{Dynamics of the universe after reheating in the $\alpha$-attractor quintessential inflation models with the exponential potential \eqref{V_exp}. We set the model parameters to  $\alpha=\frac{1}{3}$, $M=5.6\times 10^{-6}$, $\gamma=126.2$ and used the cosmological parameters today $\Omega_\Lambda=0.685$, $w_0=-1+10^{-24}$, $\Omega_\textrm{r}=9.0\times 10^{-5}$ and $H_0=5.9\times 10^{-61}$. The reheating period was indicated by the shaded region where one should not trust the numeric results, since the dynamics of the inflaton field and radiation could be affected by particle production mechanisms. (a) The evolution of effective equation of state parameter of the universe. (b) The evolution of the equation of state parameter of the inflaton field. (c) The evolution of the energy densities of inflaton field and radiation. (d) The evolution of the inflaton field for various final conditions.}\label{fig:exp-1}
\end{figure}

\begin{figure}[!htb]
	\centering
	\begin{subfigure}{0.45\textwidth}
		\includegraphics[scale=0.4]{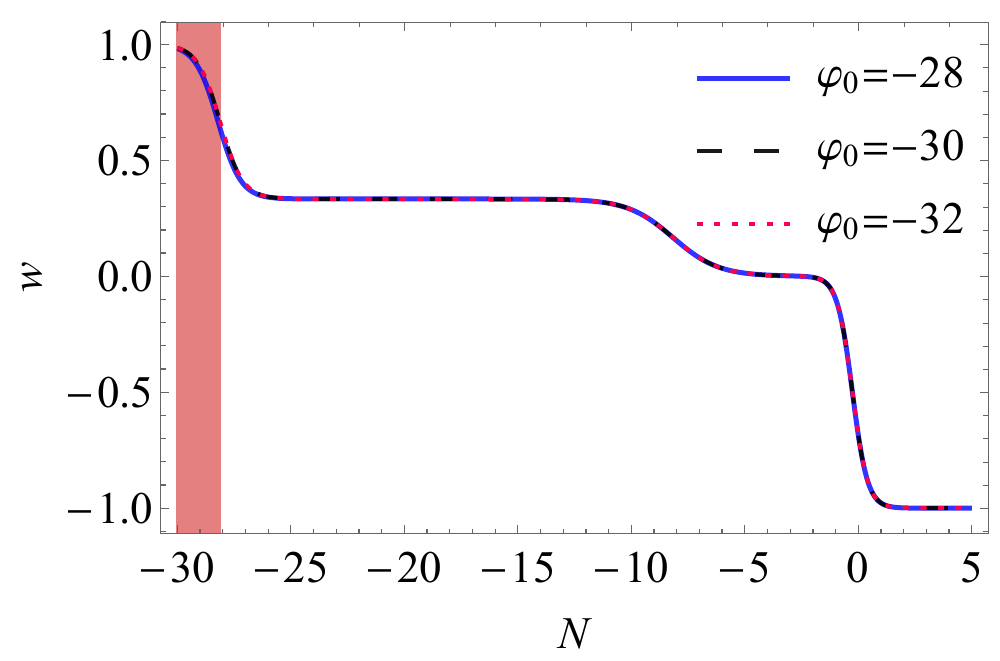}
		\subcaption{}\label{fig_exp_w-N-2}
	\end{subfigure}
	\begin{subfigure}{0.45\textwidth}
		\includegraphics[scale=0.4]{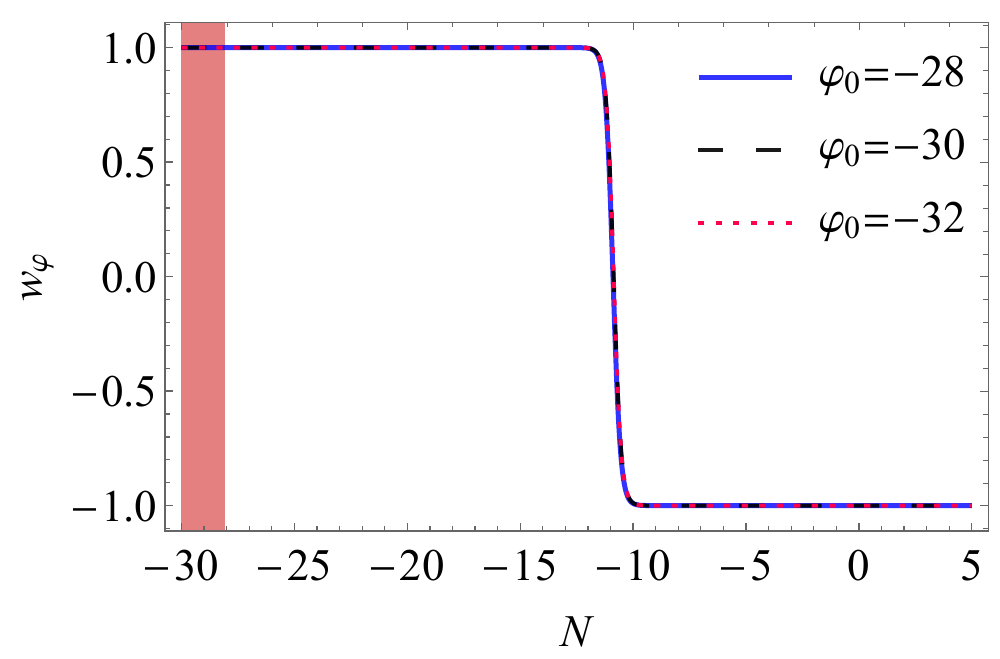}
		\subcaption{}\label{fig_exp_wphi-N-2}
	\end{subfigure}
	\begin{subfigure}{0.45\textwidth}
		\includegraphics[scale=0.4]{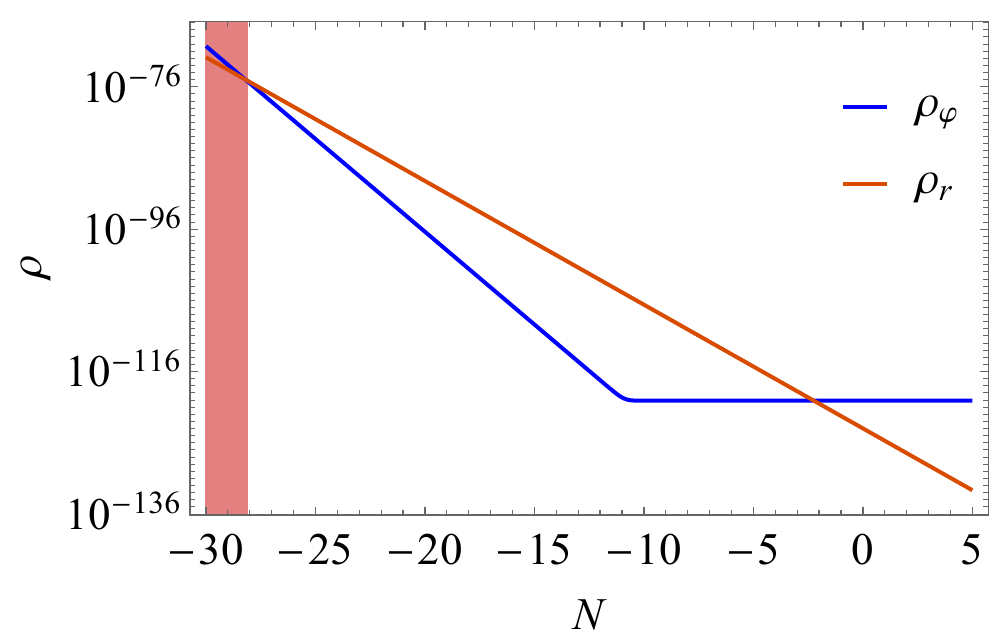}
		\subcaption{}\label{fig_exp_rho-N-2}
	\end{subfigure}
	\begin{subfigure}{0.45\textwidth}
		\includegraphics[scale=0.4]{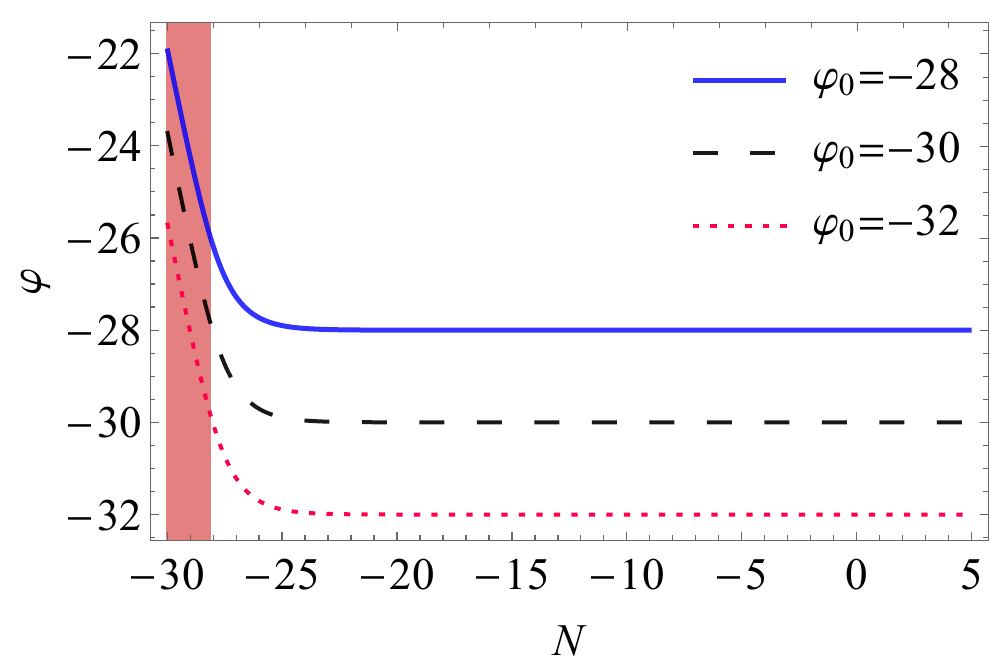}
		\subcaption{}\label{fig_exp_phi-N-2}
	\end{subfigure}
	\caption{The same as figure \ref{fig:exp-1}, but with $w_0=-1+10^{-28}$.}\label{fig:exp-2}
\end{figure}

As we mentioned earlier, the potential \eqref{V_exp} asymptotically approaches a constant value for $\varphi\ll -\sqrt{6\alpha}$, so we technically can not fix uniquely the value of the inflaton field today from eq. \eqref{phi0}. Instead, it imposes a constraint on the model parameters such that
\begin{equation}\label{M-gamma}
	M^2e^{-2\gamma}=\frac{3H_0^2\Omega_\Lambda(1-w_0)}{2}.
\end{equation}
 The requirement of negligible gradient of the potential at the present time reads
\begin{equation}\label{DVphiexp-1}
	\vert 3H_0\varphi'(0)\vert>\bigg\vert\frac{V_\varphi(\varphi_0)}{H_0}\bigg\vert,
\end{equation}
which upon substitution of eq. \eqref{V_exp} with eqs. \eqref{phi0} and \eqref{dotphi0} boils down to 
\begin{equation}\label{DVphiexp-2}
	\varphi_0 < -\sqrt{6\alpha}\cosh^{-1}\left(\sqrt{\frac{(1-w_0)\gamma\Omega_\Lambda}{3\sqrt{8(1+w_0)\alpha\Omega_\Lambda}}}\right).
\end{equation}
We choose the model parameters as $\alpha=\frac{1}{3}$, $M=5.6\times 10^{-6}$, $\gamma=126.2$ and numerically solve the equations \eqref{KG_N}--\eqref{rho-m_N} with various values of $\varphi_0$ obeying the constraint \eqref{DVphiexp-2}. The results are shown in figures \ref{fig:exp-1} and  \ref{fig:exp-2}.
We recognize that these figures display the same features as figures \ref{fig:linear-1} and  \ref{fig:linear-2}. This is not surprising as we have chosen the final conditions in such a way that the gradient of potential is negligible today, which further makes the effect of potential on the dynamics of the inflaton negligible all the way from the end of reheating to the present time. 

In the same way as the linear model, we can determine the scale factor at the end of reheating and the reheating temperature  by using the numerical result. 
The $w_0$ dependent predictions on the reheating parameters for the exponential model \eqref{V_exp} with  $\alpha=\frac{1}{3}$, $M=5.6\times 10^{-6}$ and $\gamma=126.2$ are summarized in Table \ref{Table-exp}, which again are in good agreement with the analytic relations \eqref{are-app} and \eqref{Tre-app}.
\begin{table*}[htp!]
	\centering
	\caption{$w_0$-dependent predictions on reheating parameters in the quintessential $\alpha$-attractor model with the potential \eqref{V_exp}.}\label{Table-exp}
	\begin{tabular}{|c|c|c|}
		\hline
		$w_0$ & $a_\textrm{re}$  & $T_\textrm{re}$ (GeV)  \\\hline
		$-1+10^{-24}$ & $6.2\times 10^{-11}$ &  $0.0013$  \\\hline
		$-1+10^{-28}$ & $6.2\times 10^{-13}$ &  $0.13$  \\\hline
			
		\end{tabular}
\end{table*}

\subsection{$\alpha$-attractor model with exponential two-shoulder potential}
The exponential two-shoulder potential proposed in \cite{Carrasco:2015rva} reads
\begin{equation}\label{V_TS}
V(\varphi)=M^2e^{-2\gamma}\left(e^{\gamma\tanh\frac{\varphi}{\sqrt{6\alpha}}}-1\right)^2.
\end{equation}
Here $M$ is a parameter which controls the inflationary energy scale, while $\gamma$ is responsible for large hierarchy in energy scales between inflation and late-time accelerated expansion and therefore normally taken to be very large, e.g. $\gamma\sim 126$. The inflationary predictions on $n_s$ and $r$ are most sensitive to parameter $\alpha$. Inflation can be realized for $\varphi\gg\sqrt{6\alpha}$ and the late-time accelerated expansion for $\varphi\ll -\sqrt{6\alpha}$ as in the previously considered models.

As pointed out in \cite{Akrami:2017cir}, the potential gradient is strongly suppressed compared to the Hubble friction due to the exponential prefactor for $\varphi\ll -\sqrt{6\alpha}$.  Indeed, the potential \eqref{V_TS} can be approximated  for large negative $\varphi$ as
\begin{equation}\label{Vq_TS}
V(\varphi)\simeq M^2e^{-2\gamma}\left(1-4\gamma e^{-\gamma}e^{\sqrt{\frac{2}{3\alpha}}\varphi}\right),
\end{equation}
which has the gradient
\begin{equation}\label{DV_TS}
V_\varphi=-\sqrt{\frac{32}{3\alpha}}\gamma M^2e^{-3\gamma}e^{\sqrt{\frac{2}{3\alpha}}\varphi}\simeq -\sqrt{\frac{32}{3\alpha}}\gamma e^{-\gamma}e^{\sqrt{\frac{2}{3\alpha}}\varphi}V(\varphi).
\end{equation}
One can evaluate the gradient of the potential at the present time from eq. \eqref{DV_TS} by substituting eq. \eqref{phi0}
\begin{align}
V_\varphi(\varphi_0)&=-\sqrt{\frac{24}{\alpha}}\gamma {\Omega_\Lambda H_0^2(1-w_0)} e^{-\gamma}e^{\sqrt{\frac{2}{3\alpha}}\varphi_0}\nonumber\\
&\simeq -H_0^2\varphi'(0) \sqrt{\frac{32\Omega_\Lambda}{\alpha(1+w_0)}}\gamma e^{-\gamma}e^{\sqrt{\frac{2}{3\alpha}}\varphi_0}.\label{DV0_TS}
\end{align}
In the second line we used the final condition \eqref{dotphi0}.
We see from eq. \eqref{DV0_TS} that the potential gradient is negligible compared to the Hubble friction at the present time for $\gamma\sim 126$, $\varphi_0\ll -\sqrt{6\alpha}$ and $w_0$ constrained by \eqref{w0_UB} and \eqref{w0_LB}.

Inflationary predictions of this potential can be calculated numerically. We choose $\alpha=1/3$, $M=5.6\times 10^{-6}$ and $\gamma=126.23$ so as to yield $n_s=0.968$, $r=10^{-3}$ and $V(\varphi_0)\simeq M^2e^{-2\gamma}=3\Omega_\Lambda H_0^2$. 

\begin{figure}[!htb]
	\centering
	\begin{subfigure}{0.45\textwidth}
		\includegraphics[scale=0.32]{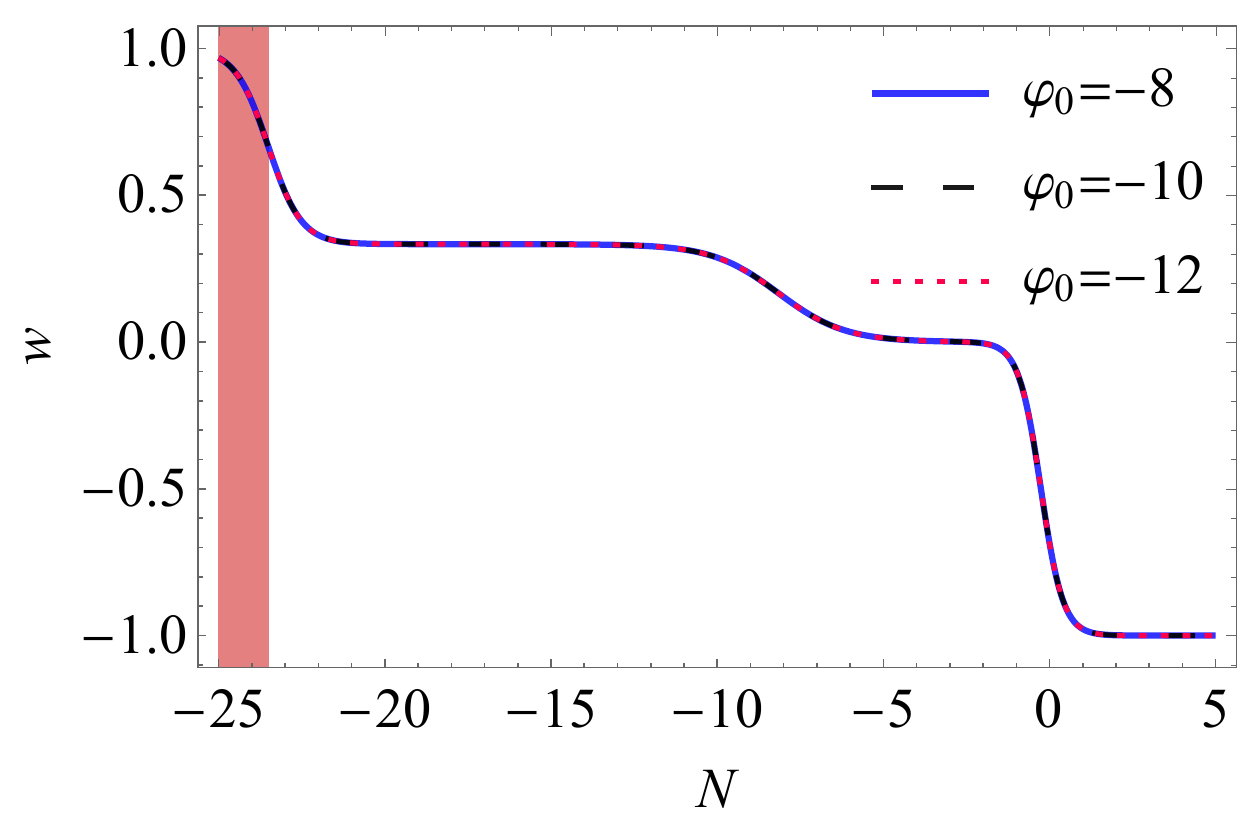}
		\subcaption{}\label{fig_TS_w-N-1}
	\end{subfigure}
	\begin{subfigure}{0.45\textwidth}
		\includegraphics[scale=0.32]{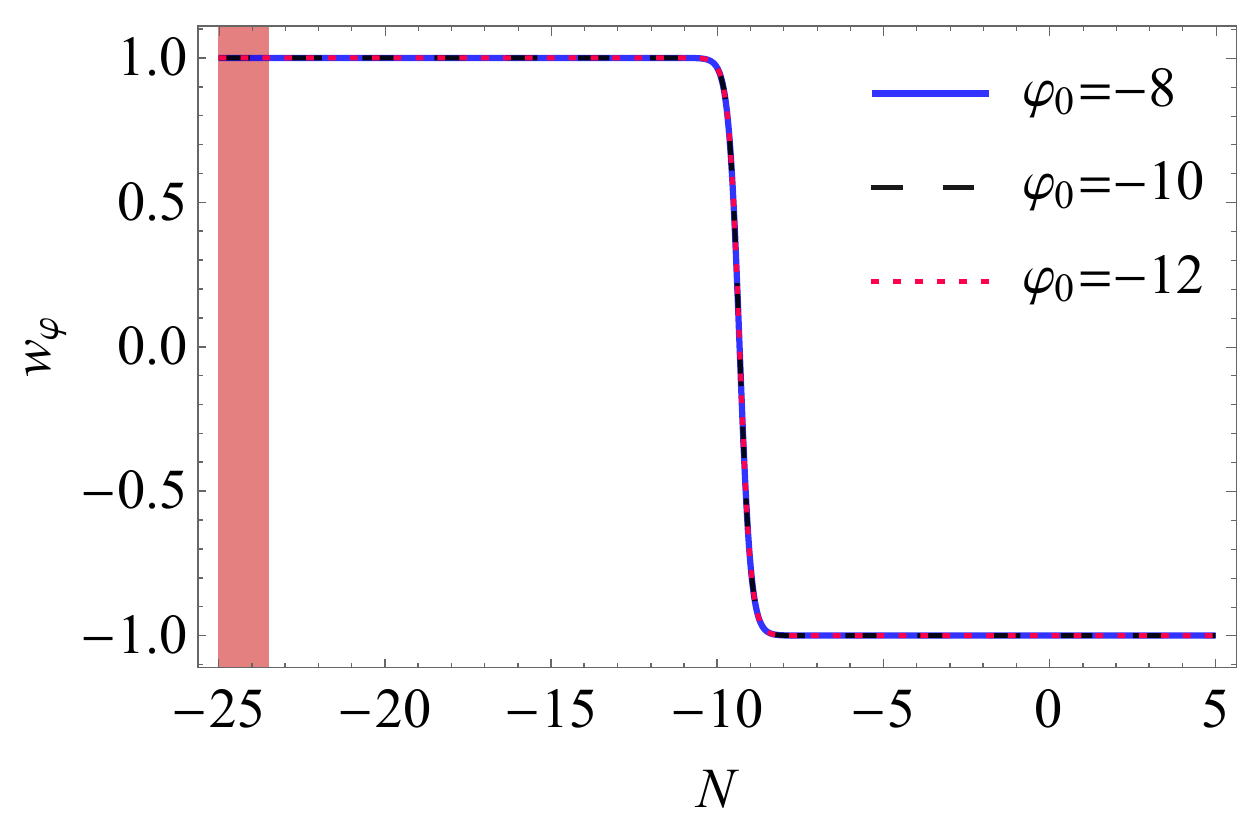}
		\subcaption{}\label{fig_TS_wphi-N-1}
	\end{subfigure}
	\begin{subfigure}{0.45\textwidth}
		\includegraphics[scale=0.32]{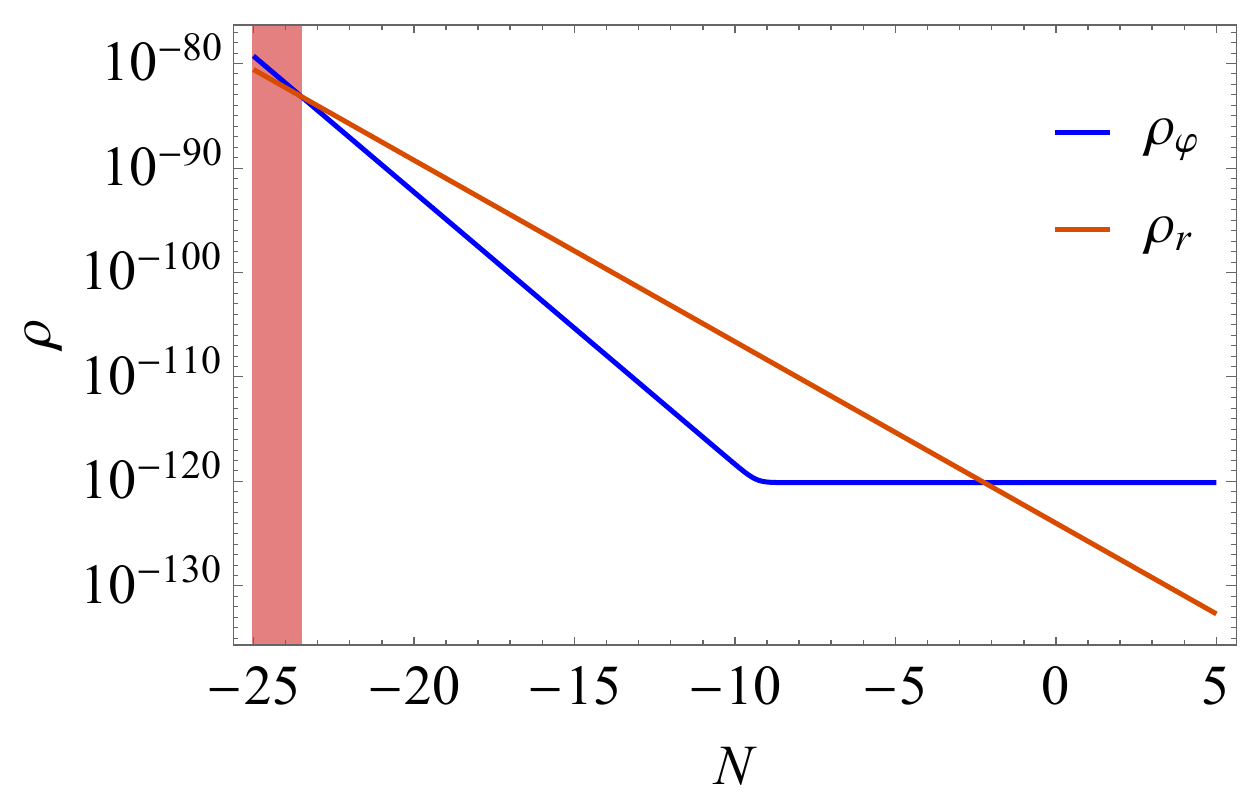}
		\subcaption{}\label{fig_TS_rho-N-1}
	\end{subfigure}
	\begin{subfigure}{0.45\textwidth}
		\includegraphics[scale=0.32]{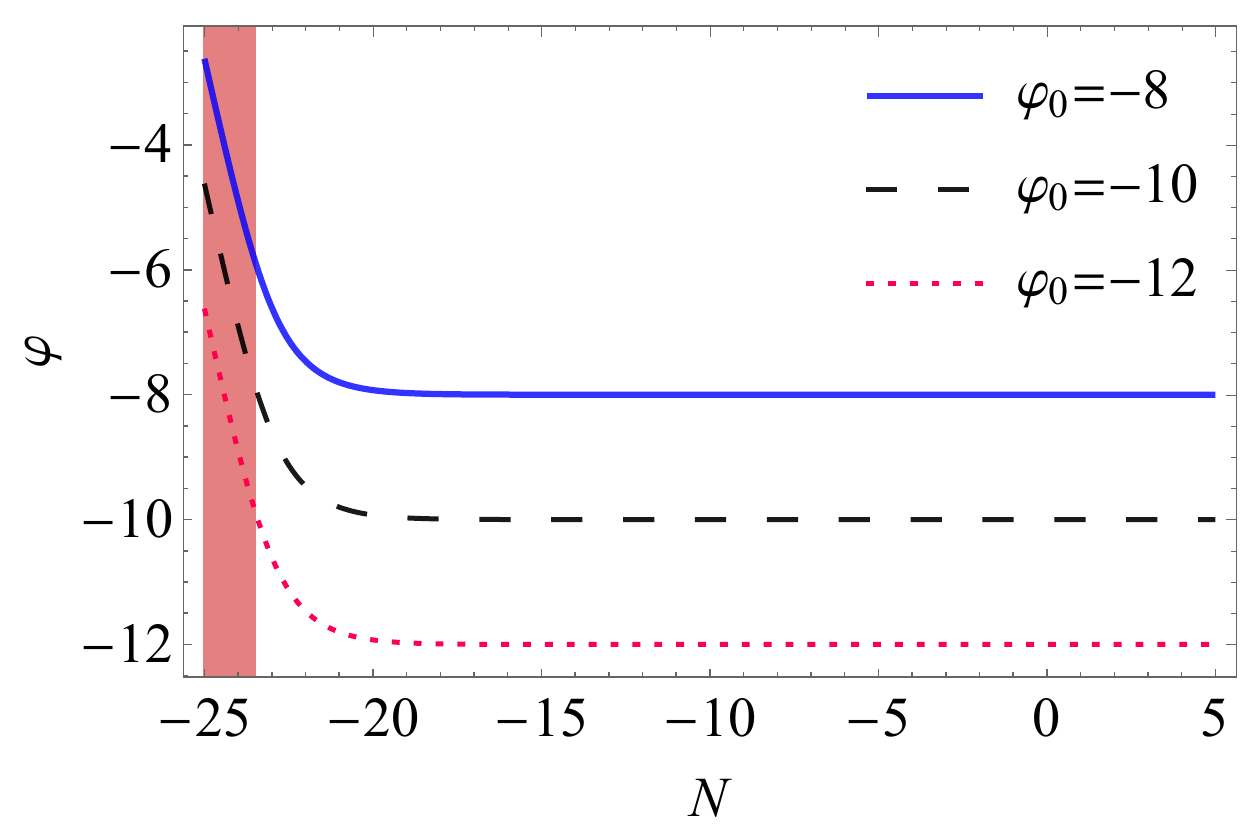}
		\subcaption{}\label{fig_TS_phi-N-1}
	\end{subfigure}
	\caption{Dynamics of the universe after reheating in the exponential two-shoulder model \eqref{V_TS}. We set the model parameters to  $\alpha=\frac{1}{3}$, $M=5.6\times 10^{-6}$, $\gamma=126.23$ and used the cosmological parameters today $\Omega_\Lambda=0.685$, $w_0=-1+10^{-24}$, $\Omega_\textrm{r}=9.0\times 10^{-5}$ and $H_0=5.9\times 10^{-61}$. The reheating period was indicated by the shaded region where one should not trust the numeric results, since the dynamics of the inflaton field and radiation could be affected by particle production mechanisms. (a) The evolution of effective equation of state parameter of the universe. (b) The evolution of the equation of state parameter of the inflaton field. (c) The evolution of the energy densities of inflaton field and radiation. (d) The evolution of the inflaton field for various final conditions.}\label{fig:TS-1}
\end{figure}

\begin{figure}[!htb]
	\centering
	\begin{subfigure}{0.45\textwidth}
		\includegraphics[scale=0.32]{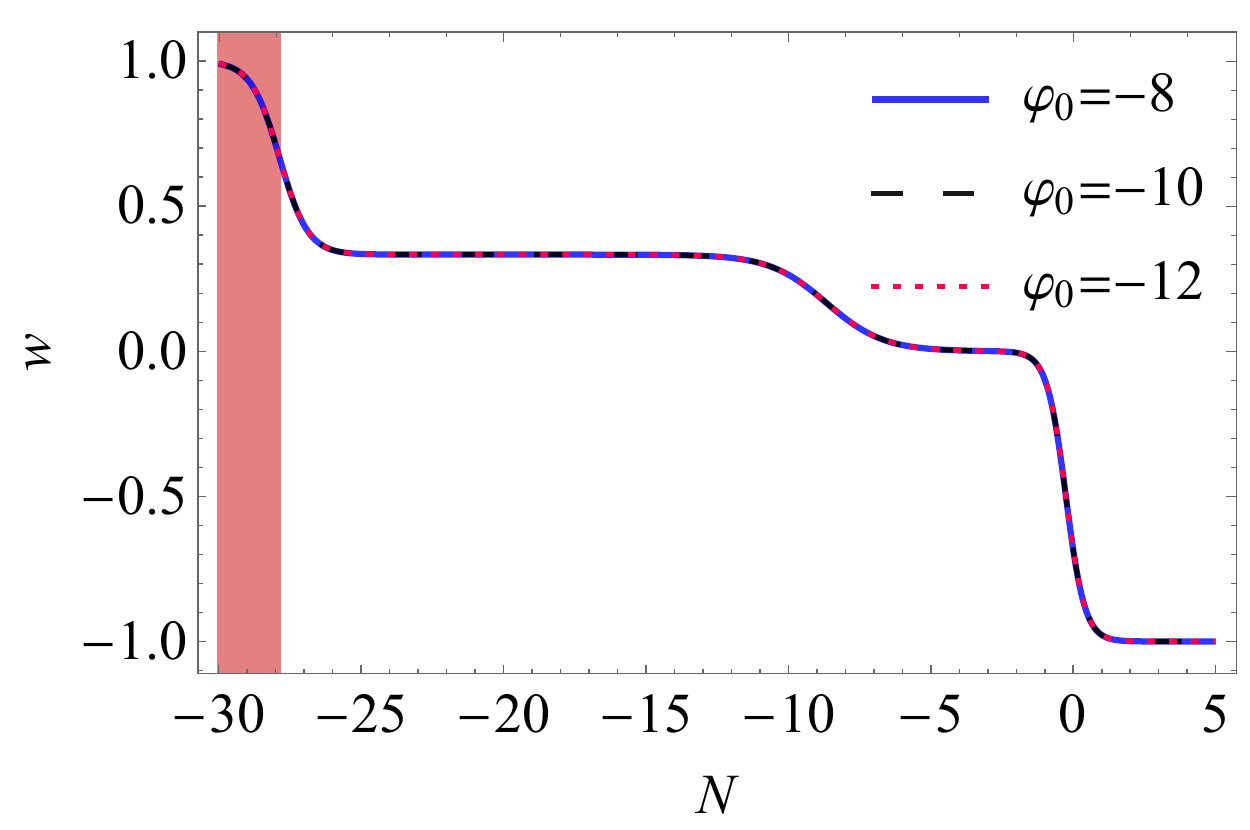}
		\subcaption{}\label{fig_TS_w-N-2}
	\end{subfigure}
	\begin{subfigure}{0.45\textwidth}
		\includegraphics[scale=0.32]{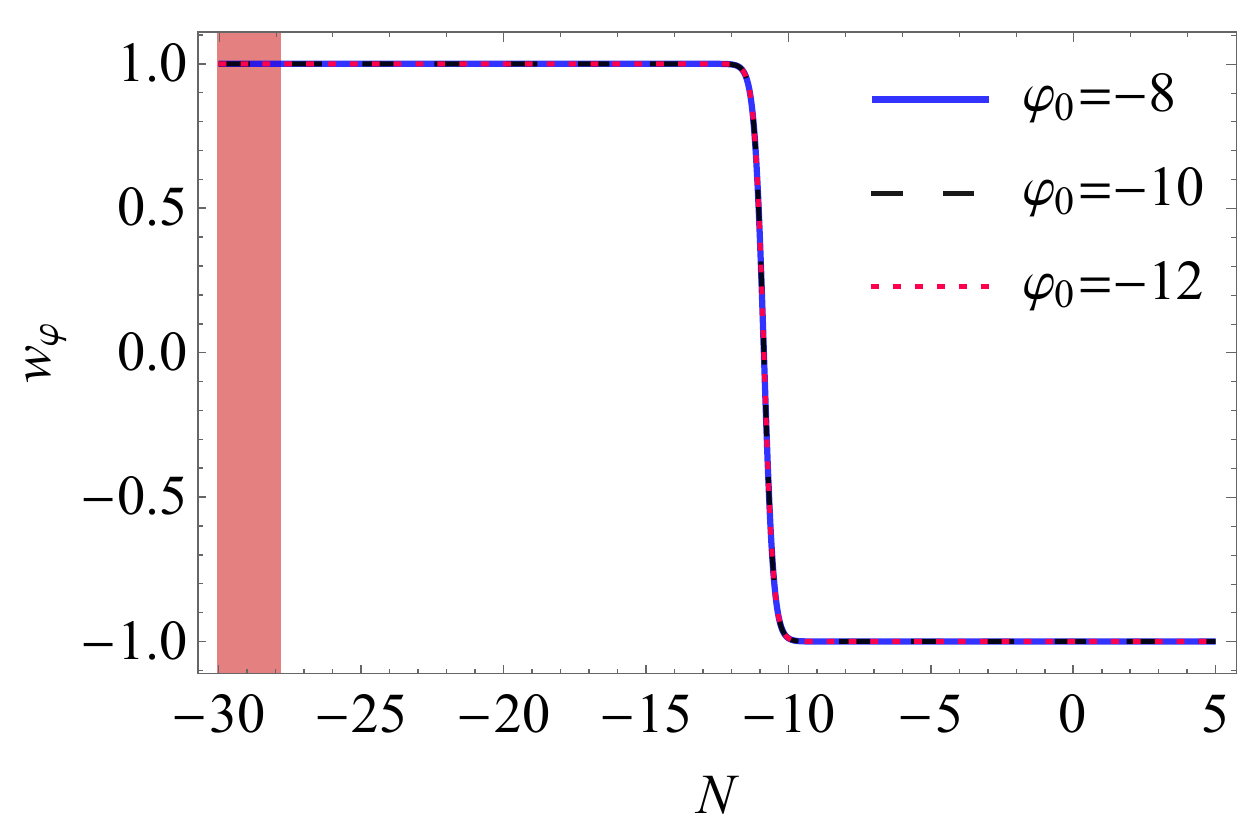}
		\subcaption{}\label{fig_TS_wphi-N-2}
	\end{subfigure}
	\begin{subfigure}{0.45\textwidth}
		\includegraphics[scale=0.32]{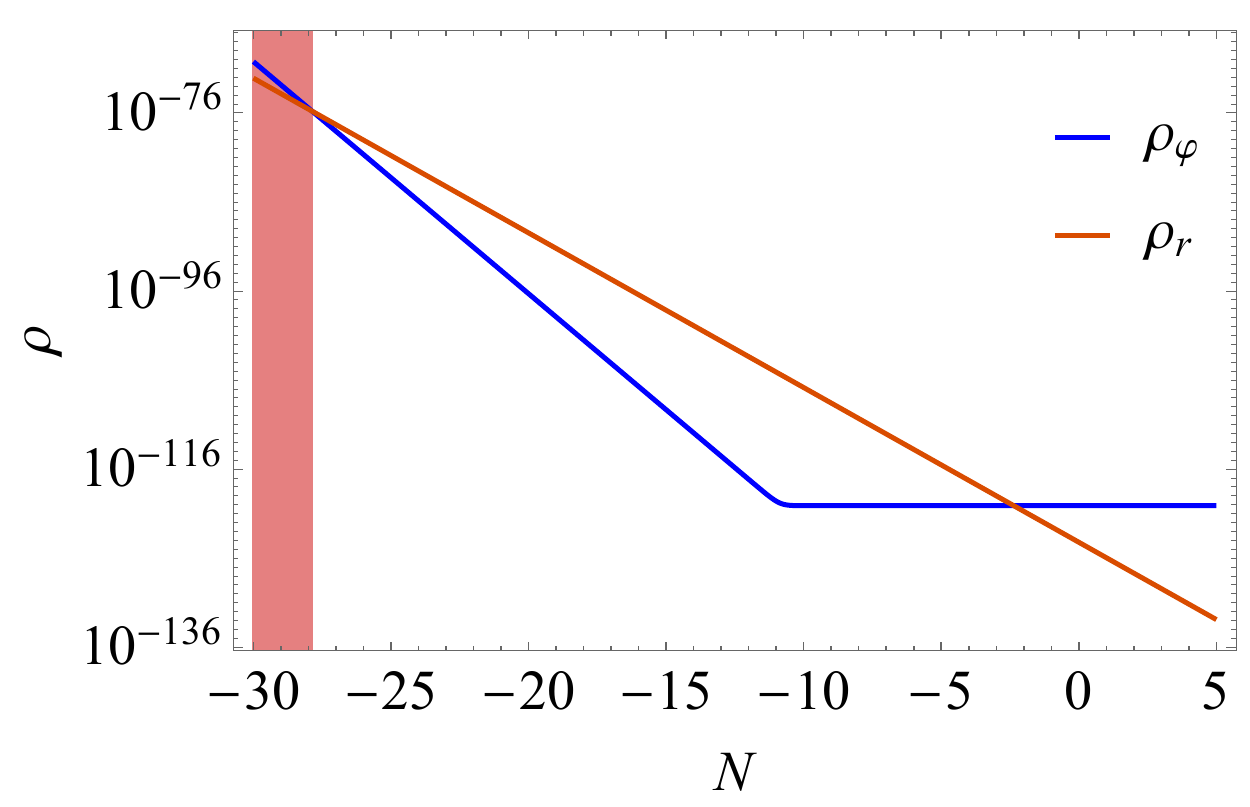}
		\subcaption{}\label{fig_TS_rho-N-2}
	\end{subfigure}
	\begin{subfigure}{0.45\textwidth}
		\includegraphics[scale=0.32]{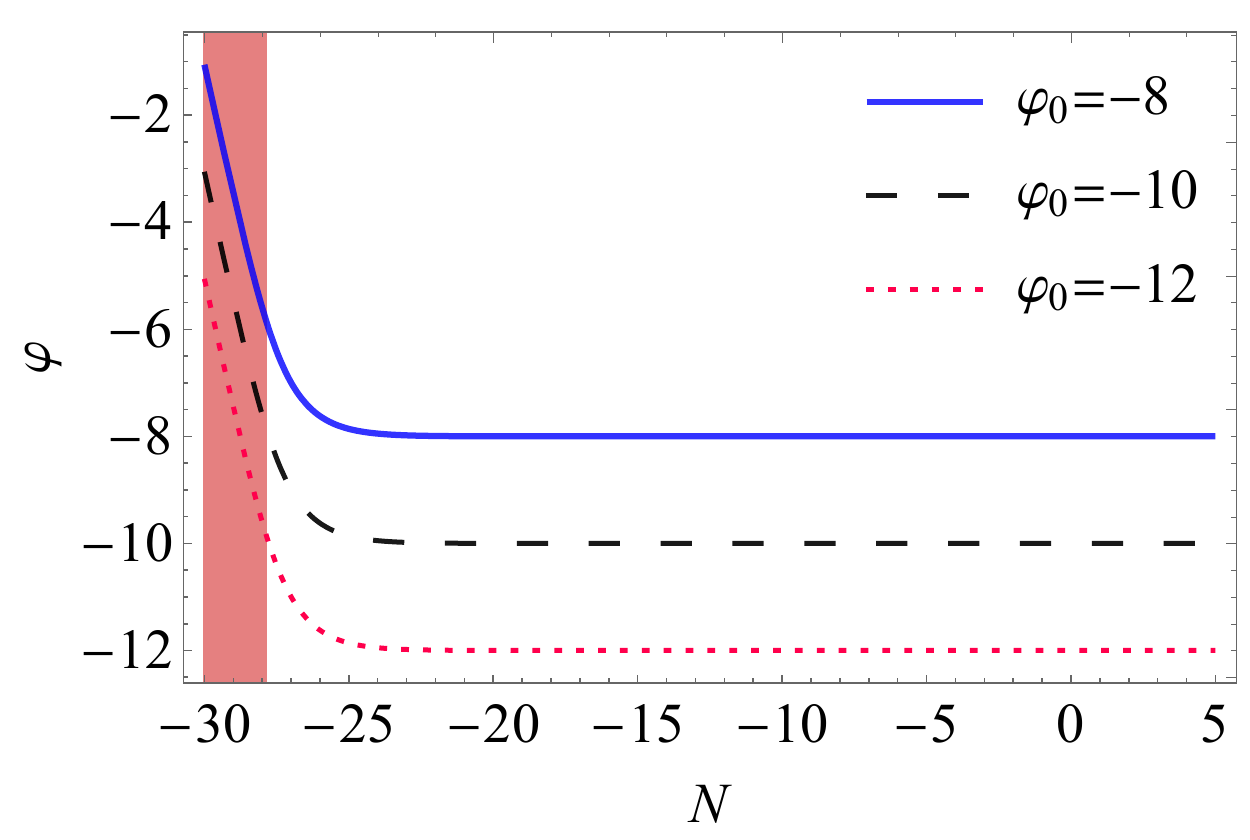}
		\subcaption{}\label{fig_TS_phi-N-2}
	\end{subfigure}
	\caption{The same as figure \ref{fig:TS-1}, but with $w_0=-1+10^{-28}$.}\label{fig:TS-2}
\end{figure}

Now we numerically solve equations \eqref{KG_N}--\eqref{rho-m_N} for different final conditions on $\varphi_0$ and $\varphi'(0)$ obeying \eqref{phi0} and \eqref{dotphi0}. The results are plotted in figures \ref{fig:TS-1} and  \ref{fig:TS-2}. These figures basically display the same behavior as those in the previous models. The values of inflaton field today are relatively small compared to the previously considered $\alpha$-attractors  due to the exponential suppression of the potential gradient. One can check that numerically determined reheating parameters are in good agreement with analytic results.

\subsection{Lorentzian quintessential inflation}
Our next example is the Lorentzian quintessential inflation model  
\cite{Benisty:2020xqm,Benisty:2020qta,Haro:2024,AresteSalo:2021lmp} which has the potential
\begin{equation}\label{V_Lorentz}
V(\varphi)=	\lambda \exp[\frac{2\xi}{\pi}\arctan(\sinh(\gamma\varphi))],
\end{equation}
where $\lambda$, $\xi$ and $\gamma$ are model parameters. 

\subsubsection{Inflationary predictions}
We first write the value of the inflaton field at the horizon crossing in terms of model parameters and the amplitude of the scalar perturbations by using eq. \eqref{Hk-phik}
\begin{equation}\label{phik_TS}
	\varphi_k=\gamma^{-1}\cosh^{-1}\left(4\gamma\xi\sqrt{\frac{3A_se^{-\xi}}{\lambda}} \right).
\end{equation}
The value of the inflaton field at the end of inflation follows from eq. \eqref{phi_end}
\begin{equation}\label{phiend_TS}
	\varphi_\textrm{end}=\gamma^{-1}\cosh^{-1}\left(\frac{\sqrt{2}\gamma\xi}{\pi}\right).
\end{equation}
Plugging eqs. \eqref{phik_TS} and \eqref{phiend_TS} into eq. \eqref{N_k}, we obtain the number of e-folds during infation as
\begin{equation}
	N_k\simeq\frac{2\pi}{\gamma}\sqrt{\frac{3A_se^{-\xi}}{\lambda}}.
\end{equation}
The scalar spectral index can be expressed as
\begin{equation}
	n_s\simeq 1 -\frac{\gamma}{\pi}\sqrt{\frac{\lambda e^\xi}{3A_s}}.\label{ns-TS}
\end{equation}
We see from eq. \eqref{ns-TS} that the Planck result $0.9607<n_s<0.9691$ \cite{Planck:2018jri} at $1\sigma$ confidence level gives the constraint
\begin{equation}
5.94\times 10^{-11}<\lambda\gamma^2 e^\xi<9.6\times 10^{-11}. 	
\end{equation}

\subsubsection{Inflaton dynamics after reheating}\label{numerics-Lorentz}
Similarly to the exponential potential \eqref{V_exp}, the asymptotic behavior of the potential \eqref{V_TS} for $\gamma\varphi\ll -1$ together with the final condition \eqref{phi0} imposes a constraint on model parameters
\begin{equation}
	\lambda e^{-\xi}=\frac{3H_0^2\Omega_\Lambda(1-w_0)}{2}.
\end{equation}
Notice that the potential gradient given by
\begin{equation}
	V_\varphi=\frac{2\gamma\xi\sech(\gamma\varphi)}{\pi}V(\varphi)
\end{equation}
 is exponentially suppressed compared to the potential for large negative value of $\gamma\varphi$. Using final conditions \eqref{phi0} and \eqref{dotphi0} we see that the potential gradient today is negligible compared to the Hubble friction for
\begin{equation}\label{phi0_TS}
\varphi_0\ll -\gamma^{-1}\cosh^{-1}\left[\frac{2\gamma\xi\sqrt{\Omega_\Lambda}}{\pi\sqrt{3 (1+w_0)}}\right].
\end{equation}

\begin{figure}[!htb]
	\centering
	\begin{subfigure}{0.45\textwidth}
		\includegraphics[scale=0.32]{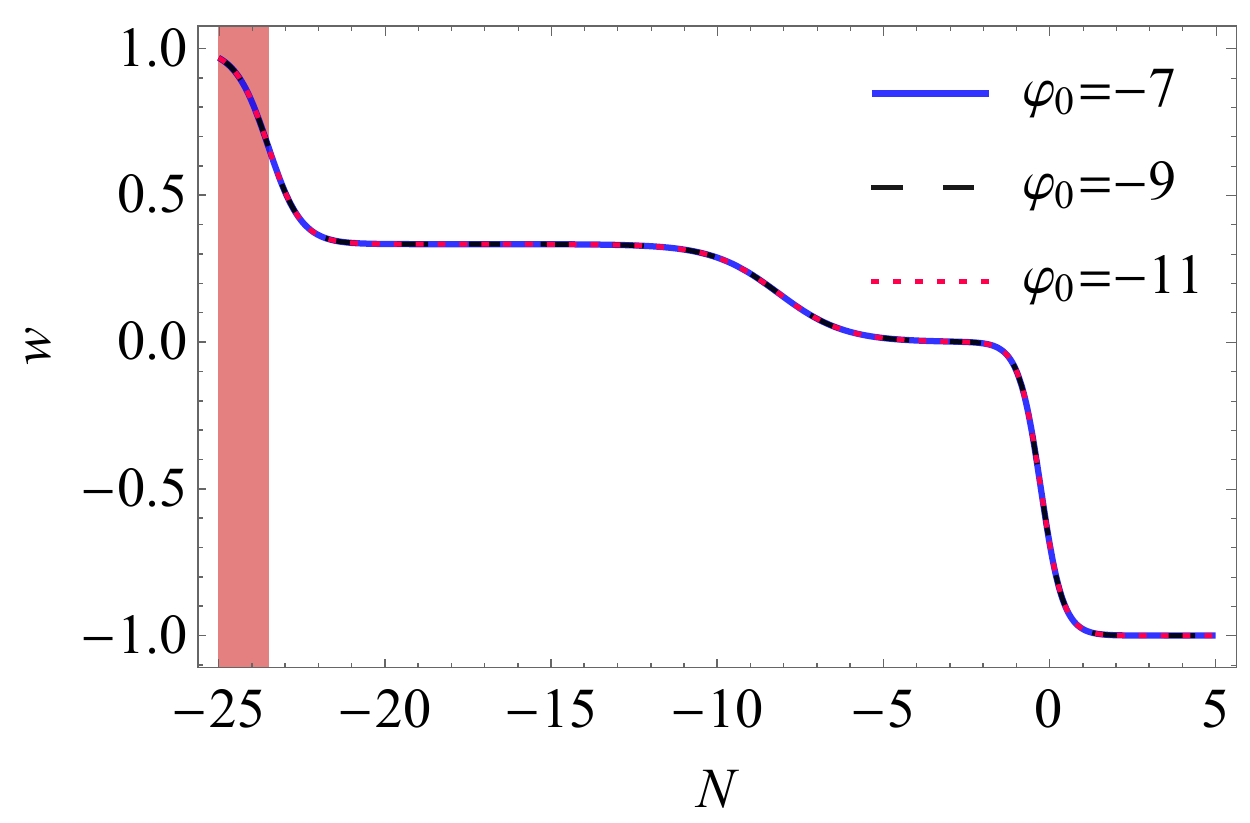}
		\subcaption{}\label{fig_L_w-N-1}
	\end{subfigure}
	\begin{subfigure}{0.45\textwidth}
		\includegraphics[scale=0.32]{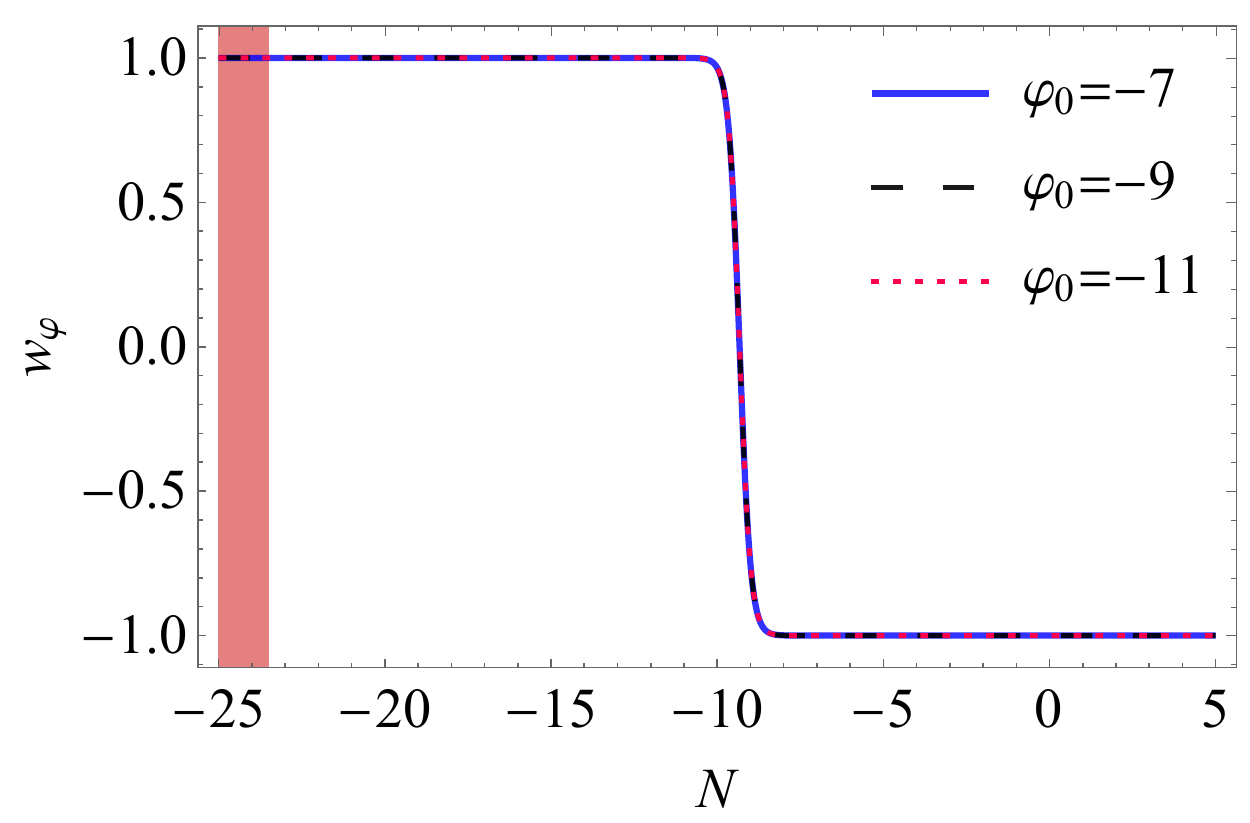}
		\subcaption{}\label{fig_L_wphi-N-1}
	\end{subfigure}
	\begin{subfigure}{0.45\textwidth}
		\includegraphics[scale=0.32]{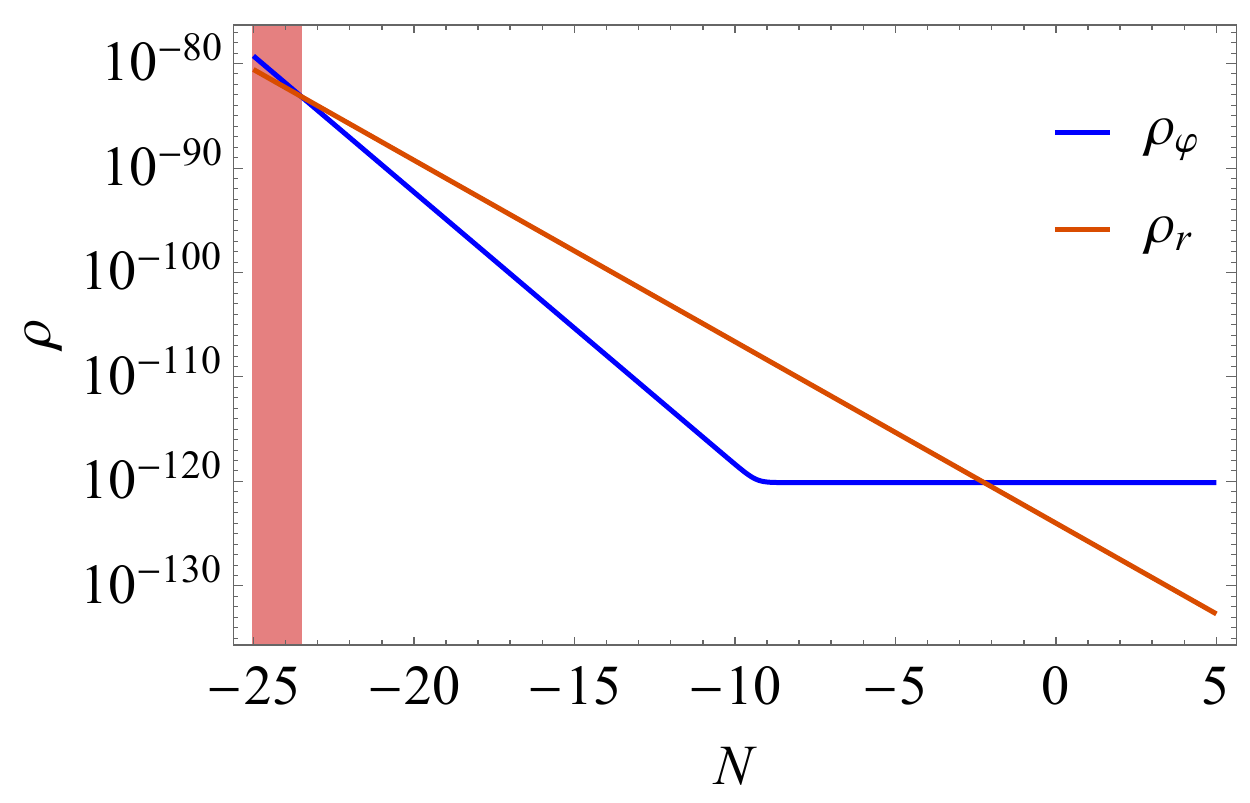}
		\subcaption{}\label{fig_L_rho-N-1}
	\end{subfigure}
	\begin{subfigure}{0.45\textwidth}
		\includegraphics[scale=0.32]{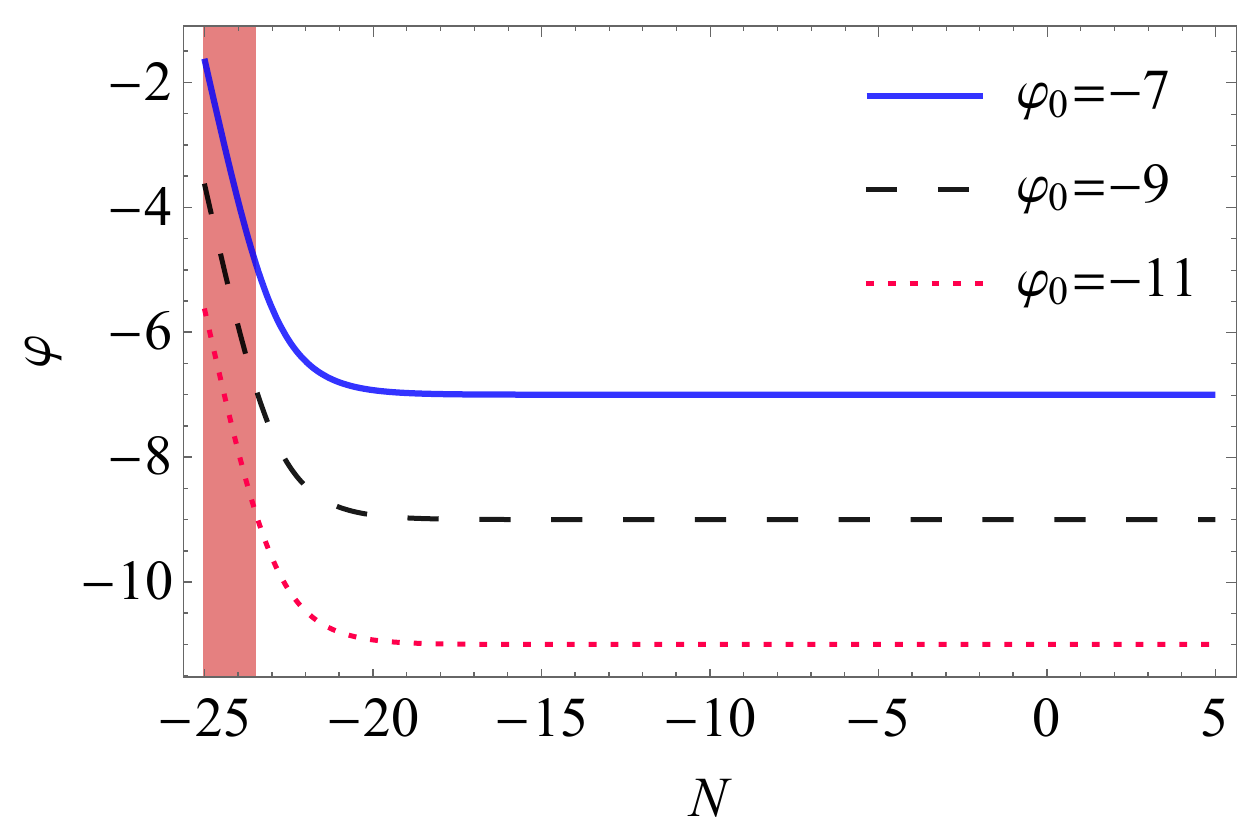}
		\subcaption{}\label{fig_L_phi-N-1}
	\end{subfigure}
	\caption{Dynamics of the universe after reheating in the quintessential Lorentzian inflation model \eqref{V_Lorentz}. We set the model parameters to  $\lambda=5.64\times 10^{-68}$, $\xi=121.8$, $\gamma=120$ and used the cosmological parameters today $\Omega_\Lambda=0.685$, $w_0=-1+10^{-24}$, $\Omega_\textrm{r}=9.0\times 10^{-5}$ and $H_0=5.9\times 10^{-61}$. The reheating period was indicated by the shaded region where one should not trust the numeric results, since the dynamics of the inflaton field and radiation could be affected by particle production mechanisms. (a) The evolution of effective equation of state parameter of the universe. (b) The evolution of the equation of state parameter of the inflaton field. (c) The evolution of the energy densities of inflaton field and radiation. (d) The evolution of the inflaton field for various final conditions.}\label{fig:Lorentz-1}
\end{figure}

\begin{figure}[!htb]
	\centering
	\begin{subfigure}{0.45\textwidth}
		\includegraphics[scale=0.32]{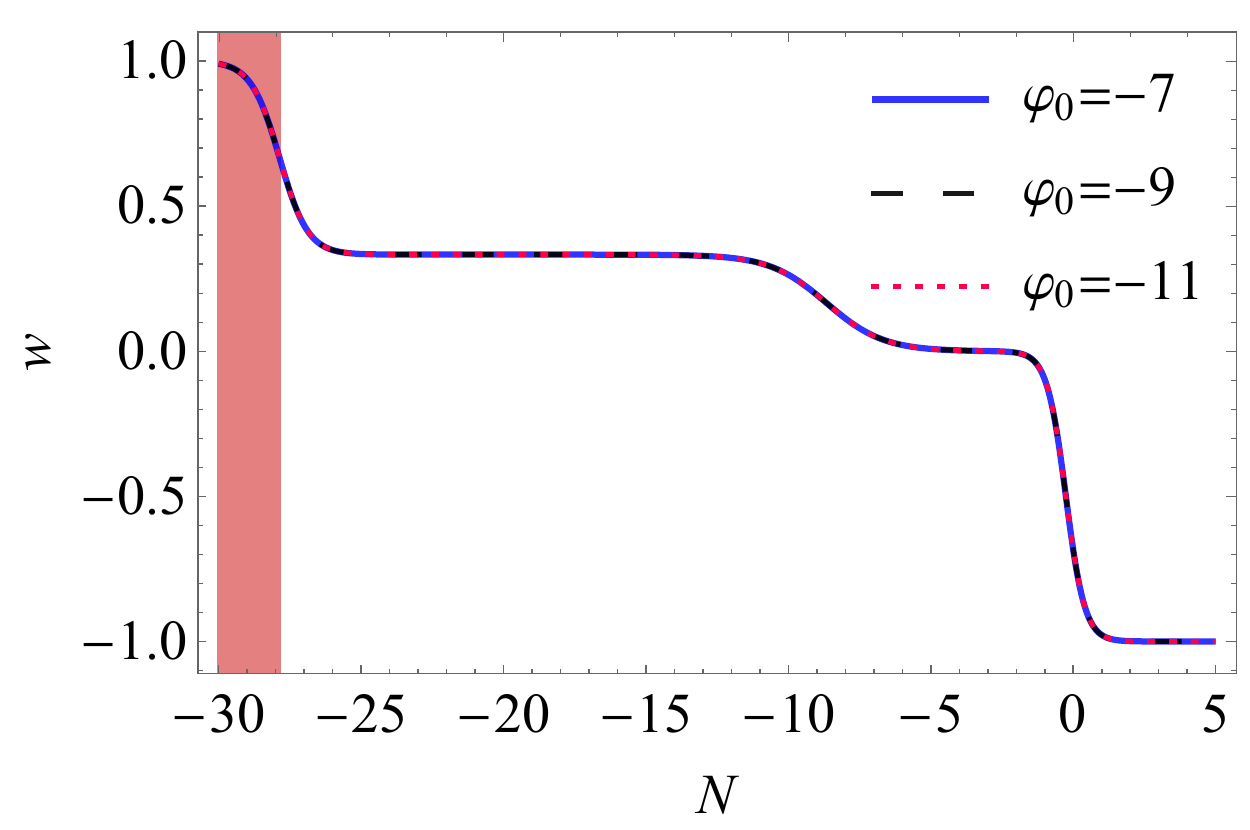}
		\subcaption{}\label{fig_L_w-N-2}
	\end{subfigure}
	\begin{subfigure}{0.45\textwidth}
		\includegraphics[scale=0.32]{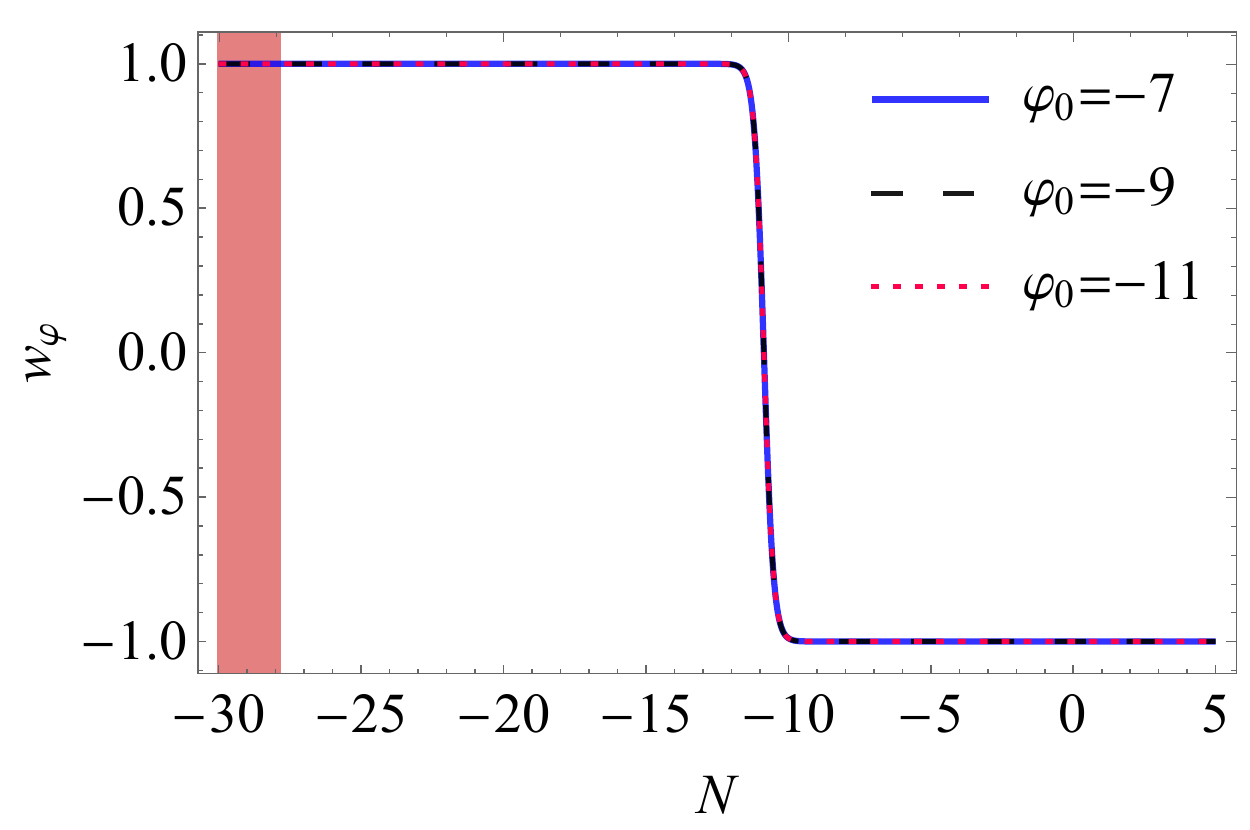}
		\subcaption{}\label{fig_L_wphi-N-2}
	\end{subfigure}
	\begin{subfigure}{0.45\textwidth}
		\includegraphics[scale=0.32]{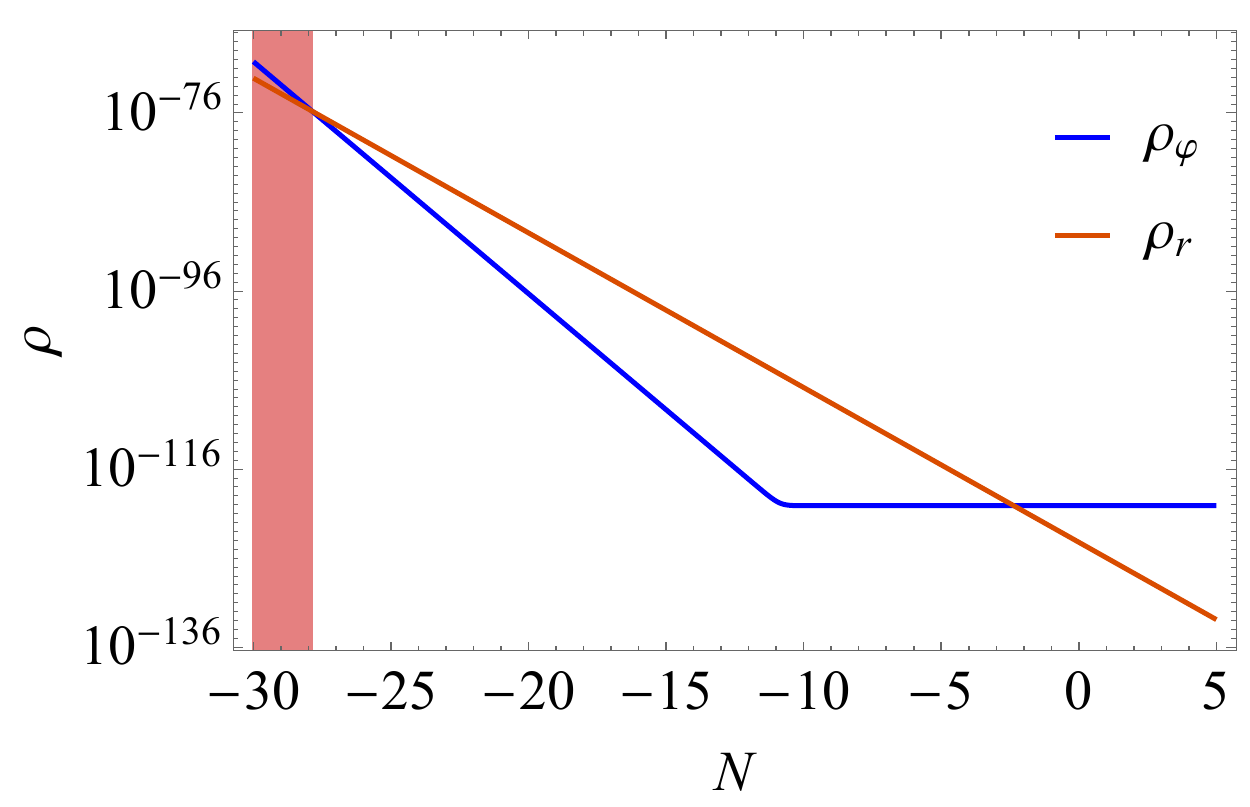}
		\subcaption{}\label{fig_L_rho-N-2}
	\end{subfigure}
	\begin{subfigure}{0.45\textwidth}
		\includegraphics[scale=0.32]{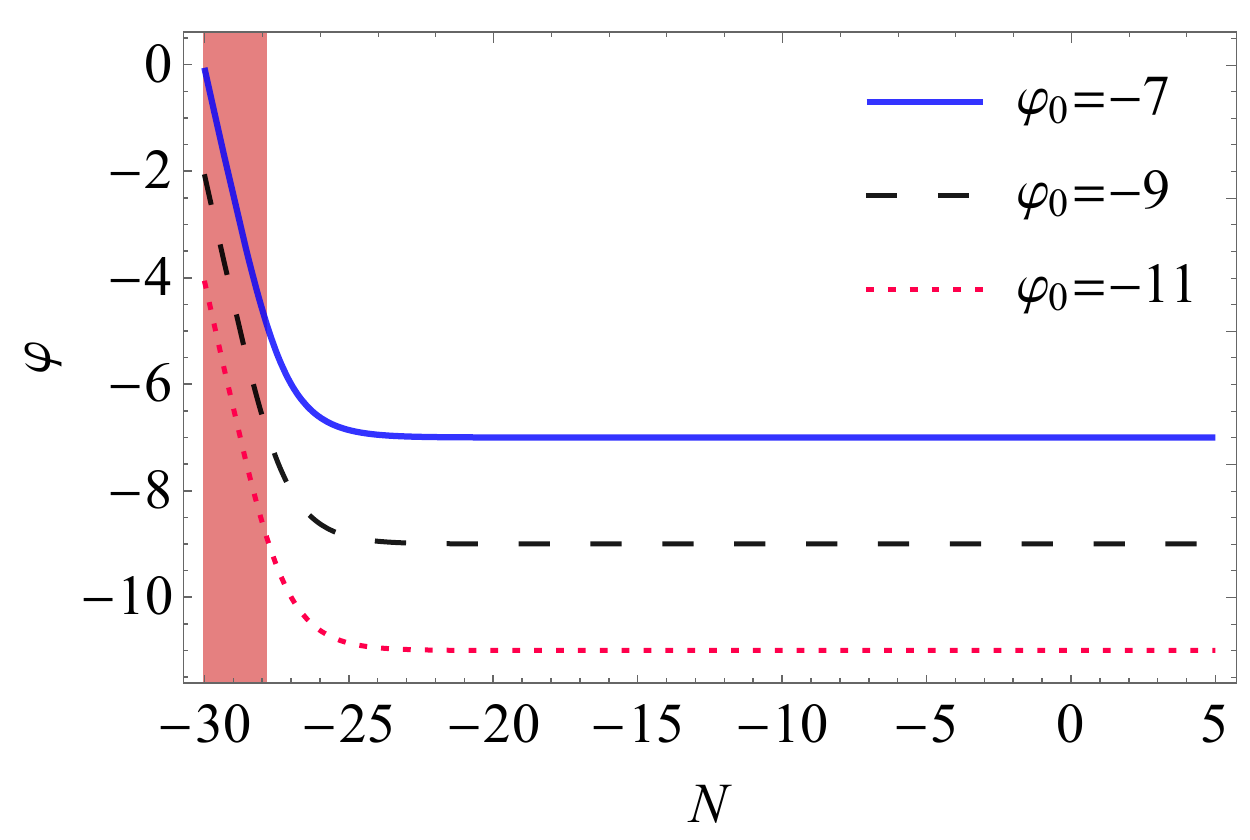}
		\subcaption{}\label{fig_L_phi-N-2}
	\end{subfigure}
	\caption{The same as figure \ref{fig:Lorentz-1}, but with $w_0=-1+10^{-28}$.}\label{fig:Lorentz-2}
\end{figure}

Now we take the model parameters as $\lambda=5.64\times 10^{-68}$, $\xi=121.8$ and $\gamma=120$ and numerically solve the equations \eqref{KG_N}--\eqref{rho-m_N} under the final conditions \eqref{phi0} and \eqref{dotphi0}. Several values of inflaton field today are chosen to obey the constraint \eqref{phi0_TS}. We depict the results in figures \ref{fig:Lorentz-1} and  \ref{fig:Lorentz-2}.  These figures also have the same behaviors as considered previously and show that the Lorentzian quintessential inflation model can result in a physically viable evolution of the universe after reheating in the regime of negligible gradient of the potential.

We numerically confirm that the predictions on the scale factor at the end of reheating and the reheating temperature are same as $\alpha$-attractor models discussed before. 


\section{Concluding remarks}\label{sec-conclusions}

In this work we proposed a new approach to build a bridge between reheating and  late-time observations in quintessential inflation. The idea is to backtrack the evolution of the inflaton field from today to the end of reheating, given the final conditions set by the observed values of the density parameter and the equation of state of dark energy at the present time (see eqs. \eqref{phi0} and \eqref{dotphi0}). 
We point out that this is in contrast to the conventional approach widely  employed in the literature, which  tracks the inflaton dynamics from early times  forward in time with the initial conditions set by hand rather than determined from observational data.

The important point to note for implementing our backtracking method is that the current value of the inflaton field can not be uniquely determined due to not only the observational uncertainties but also the asymptotic behavior of the quintessential potential. Hence we took the current value of the inflaton field as a free parameter within the range allowed by observations. Given that the potential gradient depends on the value of the inflaton field, there are two possibilities for the dynamics of the inflaton field to give rise a viable evolution of the universe after the reheating: (i) the potential gradient is negligible compared to the Hubble friction and therefore the inflaton field remains almost frozen today; (ii) the potential gradient becomes comparable to the Hubble friction and the inflaton field starts to roll very recently. In this work we assumed that the former is the case, and investigated its implications in relation with reheating. 

In particular, we derived the simple analytic relation \eqref{Tre-app} between the reheating temperature and the late-time observational parameters for dark energy. This relation is universal in the sense that it can apply to an arbitrary quintessential inflation model with any reheating mechanism. For  typical models of quintessential inflation, we numerically confirmed the validity of the analytic relation \eqref{Tre-app} and demonstrated that a physically viable evolution of the universe after reheating can be realized  when the potential gradient is negligible until today.  Several implications of eq. \eqref{Tre-app}  are placed in order:
\begin{itemize}
		\item The observations on dark energy today may be used in principle as a probe to the reheating temperature without knowledge of reheating mechanisms and potential parameters in quintessential inflation models.
	
	\item The current value of the equation of state of dark energy  in any quintessential inflation model with a successful reheating is extremely close but not equal to $-1$ (cf. eq. \eqref{w0-app}). Concretely, we found  $-1+10^{-60}\lesssim w_0 \lesssim -1+10^{-24}$ for $1\textrm{MeV}\lesssim T_\textrm{re}\lesssim 10^{15}\textrm{GeV}$ (see  \eqref{w0_UB} and \eqref{w0_LB}).
	
	\item The equation of state parameter during reheating can be related to the late-time observational parameters for a given model (see eq. \eqref{wre}). Once $w_\textrm{re}=1$ is assumed, one can further link the inflationary parameters $n_s$ and $r$ to late-time observational parameters $w_0$ and $\Omega_\Lambda$, as shown in eq. \eqref{ns-w0}.
\end{itemize}

It would be interesting to apply our backtracking method to the case in which the inflaton field unfreezes before the present time. In this case we expect that the relation between reheating and late-time observational parameters would be model dependent and complicated, as the moment when the inflaton field starts to unfreeze depends on the potential. Another interesting direction would be to extend the investigation to warm quintessential inflation \cite{Dimopoulos:2019gpz,Lima:2019yyv,Basak:2021cgk}, where the dissipation of inflaton plays an important role in the dynamics. We leave these issues for the future work.

 \bibliographystyle{JHEP3}
 \bibliography{QI_Bridge}
\end{document}